\documentclass[12pt]{article}

\usepackage{amssymb,amsmath}

\usepackage{graphicx}
\DeclareGraphicsExtensions{.pdf,.png,.jpg}

\usepackage{xcolor} 

\newcommand{\beq}[1]{\begin{equation}\label{#1}}
\newcommand{\eeq}{\end{equation}}
\newcommand{\bear}[1]{\begin{eqnarray}\label{#1}}
\newcommand{\ear}{\end{eqnarray}}

\newcommand{\const}{ {\rm const} }

\renewcommand{\theequation}{\arabic{section}.\arabic{equation}}
 \catcode`\@=11 \@addtoreset{equation}{section}\catcode`\@=12
\begin{document}

 \begin{center}
 \large \bf
The asymptotically Schwarzschild-like metric solutions

 \end{center}

 \vspace{0.3truecm}

 \begin{center}

 \normalsize\bf

 \vspace{0.3truecm}

   K. K. Ernazarov\footnote{e-mail: kubantai80@mail.ru (corresponding author)}

\vspace{0.3truecm}
  
  \it
    Institute of Gravitation and Cosmology, \\
    Peoples' Friendship University of Russia (RUDN University), \\
    6 Miklukho-Maklaya Street,  Moscow, 117198, Russian Federation \\

  \end{center}

% \begin{center}

% \normalsize\bf

% \vspace{0.3truecm}

 %  K. K. Ernazarov\footnote{e-mail: kubantai80@mail.ru (corresponding author)} \footnote{Institute of Gravitation and Cosmology, \\
 %   Peoples' Friendship University of Russia (RUDN University), \\
 %   6 Miklukho-Maklaya Street,  Moscow, 117198, Russian Federation }  

%\vspace{0.3truecm}
  
 % \it    Institute of Gravitation and Cosmology, \\
  %  Peoples' Friendship University of Russia (RUDN University), \\
  %  6 Miklukho-Maklaya Street,  Moscow, 117198, Russian Federation \\ 

 % \end{center}

\begin{abstract}
    
In this article we investigate the properties of the  asymptotically Schwarzschild-like metric as an alternative to the Schwarzschild solution in General Relativity. While asymptotically flat and similar to Schwarzschild at large distances $r$, this metric exhibits a fundamentally different strong-field behavior: it lacks an event horizon and is best interpreted as a traversable wormhole or a black hole surrounded by a specific anisotropic fluid, rather than a true vacuum solution. We analyze key phenomenological features, demonstrating significant deviations from Schwarzschild in the radii of the photon sphere and the Innermost Stable Circular Orbit (ISCO). Furthermore, we derive the Regge-Wheeler equation for gravitational and electromagnetic perturbations and compute the black hole shadow, providing a direct comparison with the Schwarzschild metric. Despite being ruled out by solar system tests, the exponential metric remains relevant for theories beyond standard GR, regular black hole models, and its connection to screened Yukawa potentials.

\end{abstract}

%\textbf{Keywords} 

%\keywords{one \and two \and three foo bar \and four $x^2$-foo}

%\keywords{Kub}

\section{Introduction}

The General Theory of Relativity, formulated by Albert Einstein in 1915, revolutionized our understanding of gravity by reimagining it not as a force, but as a manifestation of the curvature of spacetime. The fundamental equation of this theory, the Einstein Field Equations (EFE), is a complex set of nonlinear tensorial equations: $G_{\mu \nu} = 8\pi T_{\mu \nu}$. While the theory is profound, finding exact solutions to these equations for physically meaningful scenarios is notoriously difficult. Merely months after Einstein published his final theory, K. Schwarzschild provided the first and most foundational of these exact solutions - the Schwarzschild metric. This solution describes the gravitational field outside a static, spherically symmetric, non-rotating mass and remains the cornerstone for our understanding of black holes and classical tests of general relativity. K. Schwarzschild derived his solution for the metric tensor $g_{\mu \nu}$ surrounding a point mass. His achievement was remarkable, as he solved Einstein's highly complex equations under a specific set of symmetries. The key assumptions are staticity (the metric does not change with time) and spherical symmetry. This reduces the problem's complexity, allowing one to solve for the unknown functions in the line element.

As Misner, Thorne, and Wheeler (1973) elaborate in their seminal text \cite{MTW_1}, this metric satisfies the vacuum Einstein Field Equations ($R_{\mu \nu} = 0$) everywhere outside the central mass. The elegance of the solution lies in its dependence on only one parameter: the mass $M$. This simplicity belies the profound physical phenomena it predicts.

The Schwarzschild metric reveals two critical radii where the components of the metric diverge or behave strangely, leading to a long history of physical and mathematical interpretation.

The first, at $r = \frac{2GM}{c^2}$, known as the Schwarzschild radius, is a coordinate singularity. This is not a true physical singularity but an artifact of the chosen coordinate system. As Eddington \cite{Edd} first hinted and later work by Lemaitre \cite{Lem} and Synge \cite{Syn} clarified, this radius represents the event horizon of a Schwarzschild black hole. Once an object crosses this boundary, it cannot escape the gravitational pull, not even light. The event horizon is a global feature, a point of no return in the causal structure of spacetime.

The second singularity at $r = 0$ is a true curvature singularity. Unlike the coordinate singularity at the horizon, the curvature invariants, such as the Kretschmann scalar ($K = R_{\mu \nu \rho \sigma}R^{\mu \nu \rho \sigma} = \frac{48G^2M^2}{c^4r^6}$), diverge to infinity at this point \cite{Kret}. This signifies a region where the laws of general relativity break down, and a theory of quantum gravity would be required to describe the physics.

For decades, the Schwarzschild solution was considered a mathematical curiosity. However, as Oppenheimer and Snyder demonstrated in their work on gravitational collapse, it is the inevitable end-state for a sufficiently massive, non-rotating star \cite{Opp}. This work planted the seed for modern black hole astrophysics.

While the Schwarzschild metric describes a static black hole, it is the foundation upon which more realistic models are built. The Kerr metric, which describes a rotating black hole, reduces to the Schwarzschild metric when the angular momentum is zero. Today, the Schwarzschild solution is indispensable for interpreting phenomena ranging from the orbits of stars around the supermassive black hole at our Galactic Center \cite{Chez} to modeling the shadow of M87* observed by the Event Horizon Telescope Collaboration \cite{EHT}.

The asymptotically Schwarzschild-like metric is not a solution to the vacuum Einstein Field Equations $R_{\mu\nu} = 0$. If we compute the Ricci tensor $R_{\mu\nu}$ for this metric, we will find it is non-zero. This means that the such types of metrics cannot describe the empty space outside a spherically symmetric mass in standard GR. There must be some form of energy-momentum tensor $T_{\mu\nu}$ present throughout space to source this curvature. The Einstein equations 
$G_{\mu\nu} = 8\pi T_{\mu\nu}$ can be "reverse-engineered" to find the required $T_{\mu\nu}$. This effective stress-energy tensor is often interpreted as a spherical cloud of "exotic matter" or a scalar field surrounding the central mass. The properties of this required matter are non-physical (e.g., it may violate standard energy conditions), which is a major reason why the exponential metric is not considered a physically realistic model for stars or black holes.

In \cite{Brono} authors considered Einstein-Maxwell-dilaton gravity with charged dust and interaction of the form $P(\chi)F_{\mu\nu}F^{\mu\nu}$, where $P(\chi)$ is an arbitrary function of the dilaton field $\chi$ that can be normal or phantom. For any regular $P(\chi)$, static configurations are possible with arbitrary functions $g_{00} = e^{(2\gamma(x_i))}$, $i =1,2,3$ and $\chi = \chi(\gamma)$, without any assumption of spatial symmetry. The classical Majumdar-Papapetrou system is restored by putting $\chi = \text{const}$. Among possible solutions are black-hole (BH) and quasi-black-hole (QBH) ones. Some general results on BH and QBH properties are deduced and confirmed by examples. It is found, in particular, that asymptotically flat BHs and QBHs can exist with positive energy densities of matter and both scalar and electromagnetic fields.

Another regular black hole solution was introduced by I. Dymnikova, which employs an exponential function to create a de Sitter core \cite{Dym}. The model facilitates a smooth, exponential transition from a central de Sitter vacuum (where $\rho = \text{const}$) to a Schwarzschild vacuum at infinity. Derived from a spherically symmetric stress-energy tensor under a specific ansatz, this exact solution to the Einstein equations is singularity-free and reproduces the Schwarzschild geometry for large $r$ while behaving as de Sitter space for small $r$.

The "exponential metric" has circulated in the literature for approximately 60 years (\cite{Yilmaz_1} - \cite{Stanley_22}). The motivations for its study are highly varied, and its theoretical underpinnings are frequently regarded as dubious. Nevertheless, a small but persistent advocacy for this metric justifies a direct assessment of its phenomenological consequences. When analyzed on these terms, several of its properties prove to be profoundly problematic.

\section{Geometric analysis}

\subsection{Metric components}

The metric in Minkowski spacetime in spherical coordinates has the form
\begin{eqnarray}
ds^2 =- d\tilde{t}^2 + d\tilde{r}^2 + \tilde{r}^2\Big(d\theta^2 + \text{sin}^2\theta d\varphi^2\Big). \label{1A_44} 
\end{eqnarray}

The condition of spherical symmetry in a simplified form will be understood as the presence of angular coordinates only in the combination enclosed in parentheses in (\ref{1A_44}). Then, in the case of General Relativity, the metric is expected to be of the form
\begin{eqnarray}
ds^2 = -\tilde{g}_{\tilde{t}\tilde{t}}d\tilde{t}^2 + \tilde{g}_{\tilde{r}\tilde{r}}d\tilde{r}^2 + 2\tilde{g}_{\tilde{t}\tilde{r}}d\tilde{t}d\tilde{r} + f(\tilde{t}, \tilde{r})\tilde{r}^2\Big(d\theta^2 + \text{sin}^2\theta d\varphi^2\Big).
 \label{2A_44} 
\end{eqnarray}

We use the possibility of transforming two coordinates
\begin{eqnarray}
t = t(\tilde{t}, \tilde{r}), \qquad r = r(\tilde{t}, \tilde{r})
  \label{3A_44} 
\end{eqnarray}
that preserve the form of the metric (\ref{2A_44}), in such a way as to eliminate two unknown functions, reducing them to the form
\begin{equation}
 g_{tr} = 0, \qquad   f(t, r) = 1.    \label{4A_44}
\end{equation}

As a result, two unknown components of the metric tensor remain, which for the convenience of further calculations will be re-denoted as

\begin{equation}
 g_{tt} = e^{2\alpha(t, r)} , \qquad   g_{rr} = e^{2\beta(t, r)}.    \label{5A_44}
\end{equation}

Then, we use (without loss of generality) the Buchdal radial gauge obeying $\alpha(t, r) = -\beta(t, r)$, and for the \textbf{static metric} we obtain
\begin{eqnarray}
ds^2 = -e^{2\alpha(r)}dt^2 + e^{-2\alpha(r)}dr^2 + r^2\Big(d\theta^2 + \text{sin}^2\theta d\varphi^2\Big).
 \label{6A_44} 
\end{eqnarray}

The following conditions are set for the metric: 

- We assume the spacetime to be asymptotically flat, so the metric components $e^{2\alpha(r)}$ and $e^{-2\alpha(r)}$
\begin{eqnarray}
\lim_{r \to +\infty} e^{2\alpha(r)} = 1, \qquad \lim_{r \to +\infty} e^{-2\alpha(r)} = 1.
 \label{7A_44A} 
\end{eqnarray}

- For big enough value of $r$ the metric component $g_{tt}$ must have asymptotic behavior like a Schwarzschild metric component
\begin{eqnarray}
e^{2\alpha(r)} \sim \Bigg(1 - \frac{2M}{r}\Bigg) + \mathcal{O}\Big(\frac{1}{r^n}\Big).
 \label{7A_44} 
\end{eqnarray}

For the Schwarzschild metric the function $\alpha(r)$ has the form:
\begin{eqnarray}
\alpha(r) = \frac{1}{2}\text{ln}\Bigg(1 - \frac{2M}{r}\Bigg).
 \label{7A_4488} 
\end{eqnarray}

To solve Einstein’s equations, we will follow the usual procedure, which consists of the following steps:

- From certain considerations, the expected form of the covariant components of the metric tensor is specified;

- The contravariant components of the metric tensor are expressed in terms of the covariant components;

- The components of the Christoffel symbols are found in terms of the unknown functions;

- The components of the Ricci tensor are written down;

- The system of Einstein’s equations is solved.

\subsection{Christoffel symbols of the second kind and Ricci tensor}

The forty components of the Christoffel symbols, written in terms of the metric components by means of formula
\begin{eqnarray}
\Gamma_{\mu \nu}^{\lambda} = \frac{g^{\lambda \sigma}}{2}\Bigg(\frac{\partial g_{\sigma \nu}}{\partial x^{\mu}} + \frac{\partial g_{\mu\sigma}}{\partial x^{\nu}} - \frac{\partial g_{\mu \nu}}{\partial x^{\sigma}}\Bigg),
 \label{8A_44} 
\end{eqnarray}
will be calculated in four steps.

First, we will find the four components that have the same indices. They are represented in the form
\begin{eqnarray}
\Gamma_{\mu \mu}^{\mu} = \frac{g^{\mu\mu}}{2}\cdot \frac{\partial g_{\mu\mu}}{\partial x^{\mu}} 
 \label{9A_44} 
\end{eqnarray}
(no summation over repeated indices) and have the following values:
\begin{eqnarray}
\Gamma_{tt}^{t} = 0; \quad \Gamma_{rr}^{r} = -\dot{\alpha}(r); \quad  \Gamma_{\theta\theta}^{\theta} =\Gamma_{\varphi\varphi}^{\varphi} =0.
 \label{10A_44} 
\end{eqnarray}
Here and in what follows we denote $\dot{f}(r) = \frac{df(r)}{dr}$.

12 Christoffel symbol components with identical lower indices are represented in the form
\begin{eqnarray}
\Gamma_{\mu \mu}^{\sigma} =- \frac{g^{\sigma\sigma}}{2}\cdot \frac{\partial g_{\mu\mu}}{\partial x^{\sigma}}, \qquad \mu \neq \sigma
 \label{11A_44} 
\end{eqnarray}
and non zero components are found in the form
\begin{eqnarray}
\Gamma_{tt}^{r} = \dot{\alpha}(r)e^{4\alpha(r)}; \quad \Gamma_{\theta\theta}^{r} =-re^{2\alpha(r)}; \quad  \Gamma_{\varphi\varphi}^{r} =-re^{2\alpha(r)}sin^2\theta, \quad \Gamma_{\varphi\varphi}^{\theta} = -\text{sin}\theta \text{cos}\theta.
 \label{12A_44} 
\end{eqnarray}

12 Christoffel symbol components with the same one upper and one lower index are represented in the form
\begin{eqnarray}
\Gamma_{\mu \sigma}^{\sigma} = \frac{g^{\sigma\sigma}}{2}\cdot \frac{\partial g_{\sigma\sigma}}{\partial x^{\mu}}, \qquad \mu \neq \sigma
 \label{13A_44} 
\end{eqnarray}
and non zero components are found in the form
\begin{eqnarray}
\Gamma_{tr}^{t} = \Gamma_{rt}^{t} = \dot{\alpha}(r); \quad \Gamma_{r\theta}^{\theta} = \Gamma_{\theta r}^{\theta} = \frac{1}{r}; \quad \Gamma_{r\varphi}^{\varphi} = \Gamma_{\varphi r}^{\varphi} = \frac{1}{r}; 
\quad \Gamma_{\theta \varphi}^{\varphi} = \Gamma_{\varphi \theta}^{\varphi} =\text{cot}\theta.
 \label{14A_44} 
\end{eqnarray}

12 Christoffel symbol components with all different indices in this metric are equal to zero:
\begin{eqnarray}
\Gamma_{\nu \lambda}^{\mu} = 0, \qquad \mu \neq \nu \neq \lambda.
 \label{15A_44} 
\end{eqnarray}

Non zero components of Ricci tensor are:
\begin{eqnarray}
R_{tt} = e^{4\alpha(r)}\Bigg(\ddot{\alpha}(r) + 2\dot{\alpha}(r)^2 + \frac{2\dot{\alpha}(r)}{r}\Bigg), \\
R_{rr} = - \Bigg(\ddot{\alpha}(r) + 2\dot{\alpha}(r)^2 + \frac{2\dot{\alpha}(r)}{r}\Bigg),  \\
R_{\theta\theta} = 1 - e^{2\alpha(r)}\left(1+2r\dot{\alpha}(r)\right), \\
R_{\varphi\varphi} = \left[1 - e^{2\alpha(r)}\left(1+2r\dot{\alpha}(r)\right)\right]sin^2\theta,
 \label{16A_44} 
\end{eqnarray}
and Ricci scalar is equal to
\begin{eqnarray}
R = \frac{2}{r^2} - 2e^{2\alpha(r)}\Bigg(\ddot{\alpha}(r) + 2\dot{\alpha}(r)^2 + \frac{4\dot{\alpha}(r)}{r} + \frac{1}{r^2}\Bigg).
 \label{17A_44} 
\end{eqnarray}

\section{ISCO and photon sphere analysis}

Of the many exotic phenomena predicted by Einstein's theory of general relativity, few are as conceptually rich and observationally significant as the photon sphere and the innermost stable circular orbit (ISCO). These are not physical structures, but rather regions of extreme spacetime geometry around massive, compact objects like black holes. They represent fundamental boundaries that dictate the final possible states of motion for both matter and light, serving as the ultimate cosmic arenas where the classical physics of Newton gives way entirely to the warped dynamics of Einstein.

The photon sphere is a spherical region at a specific distance from a black hole where the curvature of spacetime is so severe that the paths of light itself are forced into circular orbits. For a non-rotating Schwarzschild black hole, this radius is at 1.5 times the Schwarzschild radius (the event horizon). To understand its bizarre nature, imagine a photon traveling tangentially at this exact distance. The gravitational pull is so immense that to escape, the photon would have to move radially outward at the speed of light. But in the warped geometry near the black hole, the radial direction itself is so distorted that "outward" becomes a path that merely leads into another circular orbit. These orbits are, however, profoundly unstable. Like a pencil balanced perfectly on its tip, the slightest perturbation—either inward or outward—will send the photon either spiraling catastrophically across the event horizon or flying away to infinity. This instability is what creates the black hole's "shadow," a dark silhouette surrounded by a bright, narrow ring of light from photons that are barely skimming this gravitational precipice.

Closer to the black hole, but still outside the photon sphere, lies another critical boundary: the Innermost Stable Circular Orbit (ISCO). This is the smallest possible radius at which a massive particle, like an electron or a planet, can maintain a stable circular orbit. Within Newtonian gravity, orbits are theoretically possible at any distance from a central mass. General relativity shatters this notion. As one moves closer to the black hole, the "centrifugal force" needed to maintain a circular orbit does not increase fast enough to counterbalance the escalating gravitational pull. For a Schwarzschild black hole, the ISCO resides at 3 times the Schwarzschild radius. Inside this radius, circular orbits are no longer stable; any small disturbance will cause the orbiting object to plunge inexorably inward, past the event horizon. The ISCO is therefore the true "edge of the abyss" for material objects. It is the inner rim of an accretion disk—the superheated, swirling maelstrom of gas and dust falling into a black hole. The material orbits stably until it reaches the ISCO, at which point it makes its final, fatal plunge, releasing a final burst of energy.

The relationship between the ISCO and the photon sphere is a elegant hierarchy of cosmic limits. The photon sphere represents the ultimate limit for massless particles (light), while the ISCO is the ultimate limit for massive particles (matter). The photon sphere always lies inside the ISCO, creating a layered structure: from the outside in, one first encounters the ISCO, the last stable orbit for matter; then, closer in, the photon sphere, the unstable orbit for light; and finally, the event horizon, the point of no return. This sequence underscores a key principle: the effects of strong gravity become more extreme in steps, first destabilizing matter, then capturing light, and finally severing all connection with the external universe.

Furthermore, these features are not static. For a rotating Kerr black hole, the dynamics become even more fascinating. The rotation of the black hole drags the surrounding spacetime along with it, an effect known as frame-dragging. This rotation breaks the symmetry, making the ISCO and photon sphere dependent on whether an object is orbiting with the spin (prograde) or against it (retrograde). For a maximally rotating black hole, the prograde ISCO can be as close as the event horizon itself, while the prograde photon sphere also moves inward. This dependence makes the ISCO and photon sphere not just markers of gravity's strength, but also exquisite probes of a black hole's spin—a fundamental property that can be measured by observing the X-ray emission from accretion disks.

%\begin{equation}
%\begin{gathered}
%ds^2 = -e^{2\alpha\left(r\right)}dt^2 + e^{-2\alpha\left(r\right)}dr^2 + r^2d\Omega^2  \label{7E} 
%\end{gathered}
%\end{equation}

Let us now find the photon sphere for massless particles and the ISCO for massive particles as functions of the $\alpha\left(r\right)$ in the metric (\ref{6A_44}). Consider the tangent vector to a particle's worldline, parameterized by an affine parameter $\lambda$, Lagrangian $L$ has following form:

\begin{equation}
%\begin{gathered}
L = g_{\mu\nu}\frac{dx^\mu}{d\lambda}\frac{dx^\nu}{d\lambda} = -g_{tt}\Bigg(\frac{dt}{d\lambda}\Bigg)^2 + g_{rr}\Bigg(\frac{dr}{d\lambda}\Bigg)^2 + r^2\Biggl \{\left(\frac{d\theta}{d\lambda}\right)^2 + sin^2\theta\left(\frac{d\varphi}{d\lambda}\right)^2\Biggr\}.   \label{7E2} 
%\end{gathered}
\end{equation}

We  separate (without loss of generality)  the two physically interesting cases (timelike and null) by definition:

\begin{equation}
%\begin{gathered}
k = \begin{cases}
  -1,  & \mbox{ massive particle,} i.e. \mbox{timelike worldline}\\
 0, & \mbox{massless particle } i.e. \mbox{ null worldline.}
\end{cases}, \label{7.Y} 
%\end{gathered}
\end{equation}
That is, $\frac{ds^2}{d\lambda^2} = k$. Given the spherical symmetry of the metric, we can fix $\theta = \frac{\pi}{2}$ arbitrarily and consider the resulting equatorial problem:

\begin{equation}
%\begin{gathered}
g_{\mu\nu}\frac{dx^\mu}{d\lambda}\frac{dx^\nu}{d\lambda} = -g_{tt}\Bigg(\frac{dt}{d\lambda}\Bigg)^2 + g_{rr}\Bigg(\frac{dr}{d\lambda}\Bigg)^2 + r^2\left(\frac{d\varphi}{d\lambda}\right)^2 = k.    \label{7E2444} 
%\end{gathered}
\end{equation}

The Killing symmetries in the metric (\ref{6A_44}) yield the following expressions for the conserved energy $\varepsilon$ and angular momentum $L$ per unit mass:
\begin{eqnarray}
\begin{gathered}
e^{2\alpha\left(r\right)}\Bigg(\frac{dt}{d\lambda}\Bigg) = \varepsilon; \qquad r^2\Bigg(\frac{d\varphi}{d\lambda}\Bigg) = L.
 \label{7.AB} 
\end{gathered}
\end{eqnarray}

Hence 
\begin{eqnarray}
\begin{gathered}
e^{-2\alpha\left(r\right)}\Bigg(-\varepsilon^2 + \Big(\frac{dr}{d\lambda}\Big)^2\Bigg) + \frac{L^2}{r^2} = k,
\label{8.Y44} 
\end{gathered}
\end{eqnarray}
implying
\begin{eqnarray}
\begin{gathered}
\Big(\frac{dr}{d\lambda}\Big)^2 = \varepsilon^2 + e^{2\alpha\left(r\right)}\Bigg( k - \frac{L^2}{r^2}\Bigg)  \label{8.Y_4545} 
\end{gathered}
\end{eqnarray}
This leads to the following "effective potentials" for geodesic orbits: 
\begin{eqnarray}
\begin{gathered}
U\left(r\right) = e^{2\alpha\left(r\right)}\Bigg(- k + \frac{L^2}{r^2}\Bigg). 
 \label{8.F_44} 
\end{gathered}
\end{eqnarray}

The Lagrange equation for the radial coordinate $r$ can be readily verified to be:
\begin{eqnarray}
\begin{gathered}
2\ddot{r} + \frac{dU(r)}{dr} = 0. 
 \label{8F_55K} 
\end{gathered}
\end{eqnarray}

Alternatively, this result can be derived by differentiating Eq. (\ref{8.Y_4545}) with respect to the parameter $\lambda$ and then dividing by $\dot{r}$, provided that $\dot{r} \neq 0$. In the case where $\dot{r} = 0$, the radial equation becomes
\begin{eqnarray}
\begin{gathered}
\frac{dU(r)}{dr} = 0. 
 \label{8F_66K} 
\end{gathered}
\end{eqnarray}
which does not follow from Eq. (\ref{8.Y_4545}) and must be considered separately.

For a photon orbit in a spherically symmetric spacetime, the photon sphere's location is found by solving $V_0^{'}\left(r\right) =0.$ for the massless particle case ($\epsilon =0$). That is
\begin{eqnarray}
\begin{gathered}
U_0\left(r\right) = e^{2\alpha\left(r\right)}\Big(\frac{L^2}{r^2}\Big)  \label{7E_98} 
\end{gathered}
\end{eqnarray}
gives:
\begin{equation}
%\begin{gathered}
 \dot{U_0}\left(r\right) = \frac{2L^2e^{2\alpha\left(r\right)}}{r^3}\Biggl\{\dot{\alpha}(r)r - 1\Biggr\}.  \label{6F_44} 
%\end{gathered}
 \end{equation}

For massive particles the geodesic orbit corresponds to a timelike worldline and we have the case that $\epsilon = -1$. Therefore: 

\begin{eqnarray}
\begin{gathered}
U_{-1}\left(r\right) = e^{2\alpha\left(r\right)}\Bigg( 1 + \frac{L^2}{r^2}\Bigg),
 \label{8F_87B} 
\end{gathered}
\end{eqnarray}
which leads to:

\begin{eqnarray}
\begin{gathered}
\dot{U}(r)_{-1} = 2e^{2\alpha(r)}\Bigg(\dot{\alpha}(r)\Big(1 + \frac{L^2}{r^2}\Big) - \frac{L^2}{r^3}\Bigg). \label{7EF0} 
\end{gathered}
\end{eqnarray}
Directly solving for  r is often algebraically intractable. Therefore, to find the innermost stable orbit, we assume a fixed circular orbit at a radius $r_c$. We then express the required angular momentum $L_c$ as a function of $\alpha(r)$. The minimum value of $L_c$ (or, more practically, $L_c^2$) with respect to $r_c$  identifies this innermost orbit. This approach is chosen over a direct algebraic solution for simplicity.

Hence if  $\dot{V}(r)_{-1} = 0$, we have: 
\begin{eqnarray}
\begin{gathered}
L_c^2 = \frac{\dot{\alpha}(r)r^3}{1 - \dot{\alpha}(r)r}. \label{7G89R} 
\end{gathered}
\end{eqnarray}
Solving equation $\frac{\partial L_c}{\partial r_c} =0$, we can find the innermost orbit $r_c$.

\section{Stress-energy tensor and energy conditions}

In the Newtonian universe, gravity is sourced simply by mass. Einstein’s revolutionary insight was that gravity arises from the curvature of spacetime itself, and that the source of this curvature is everything that carries energy and momentum. The stress-energy tensor, denoted as $T_{\mu \nu}$, is the mathematical object that encapsulates this source completely. It is a symmetric, $4x4$ matrix that provides a comprehensive inventory of the energy and momentum content at a point in spacetime.

Its components have direct physical interpretations:

$T_{00}$ is the energy density, the relativistic analogue of mass density.

$T_{0i}$ is the energy flux (or momentum density) in the $i$-th direction.

$T_{i0}$ is the momentum density (equivalent to energy flux due to symmetry).

$T_{ij}$ is the stress tensor, representing the $i$-th component of force acting on a unit area with its normal in the $j$-th direction. This includes pressure (diagonal components) and shear stresses (off-diagonal components).

Through Einstein's field equations, $G_{\mu \nu} = 8\pi T_{\mu \nu}$, this tensor is directly equated to the Einstein tensor $G_{\mu \nu}$, which describes the geometry of spacetime. In essence, $T_{\mu \nu}$  is the "why" behind the "how" of spacetime curvature. It is the command that matter and energy issue to the very stage upon which they perform.

Let us examine the Einstein field equations for this spacetime. Using the mixed components $G_\nu^\mu = 8\pi T_\nu^\mu$, $\rho = -T_t^t$, $p_{\parallel} = T_r^r$, $p_{\perp} = T_\theta^\theta = T_\varphi^\varphi$, this yields the following form of the stress-energy-momentum tensor:
\begin{eqnarray}
\rho = \frac{1}{8\pi}\cdot\frac{1 - \Big(1 + 2\dot{\alpha}(r)\Big)e^{2\alpha(r)}}{r^2},  \nonumber \\
p_\parallel = \frac{1}{8\pi}\cdot \frac{\Big(1 + 2\dot{\alpha}(r)\Big)e^{2\alpha(r)} - 1}{ r^2}, \nonumber \\
 p_\perp = \frac{1}{8\pi}\cdot e^{2\alpha(r)}\Big(2\dot{\alpha}(r)^2 + \ddot{\alpha}(r) + \frac{2\dot{\alpha}(r)}{r}\Big)    
\label{SET_1} 
\end{eqnarray}

\textbf{The Null Energy Condition (NEC)} is aslightly weaker condition, it requires that the energy density plus the pressure as measured by any observer with a lightlike (null) velocity is non-negative. he NEC is crucial for many key theorems in GR, including the second law of black hole thermodynamics and the focusing theorem, which dictates that gravity always acts as a focusing lens for light rays. A necessary and suffcient condition for the null energy condition (NEC) to hold is that both $\rho + p_\parallel \ge 0$ and $\rho + p_\perp \ge 0$. Since $\rho = -p_\parallel$, the former inequality is trivially satisfied. For the second condution we obtain following inequality:

\begin{equation}
\rho + p_\perp = \frac{1}{8\pi} \Bigg(e^{2\alpha(r)}\Big(2\dot{\alpha}(r)^2 + \ddot{\alpha}(r)  - \frac{1}{r}\Big) + \frac{1}{r^2}\Bigg) \ge 0.
\label{SET_2} 
\end{equation}

\textbf{The Weak Energy Condition (WEC)} states that the energy density measured by any observer with a timelike velocity should be non-negative. In simple terms, all observers should see a positive energy density. his is a fundamental expectation from everyday experience. In order to satisfy the weak energy condition (WEC) we require the NEC be satisfied, and in addition $\rho \ge 0$. Therefore, to fully satisfy weak energy condution (WEC) one more inequality is added:

\begin{equation}
\rho = \frac{1}{8\pi}\cdot\frac{1 - \Big(1 + 2\dot{\alpha}(r)\Big)e^{2\alpha(r)}}{r^2} \ge 0.
\label{SET_3} 
\end{equation}

\textbf{The Strong Energy Condition (SEC)} is important for the singularity theorems of Penrose and Hawking, states that $\rho + p_\parallel + 2p_\perp \ge 0$,  $\rho + p_\parallel  \ge 0$,  $\rho + p_\perp \ge 0$. It essentially ensures that gravity is always attractive. Notably, a cosmological constant with $p = -\rho$ violates the SEC, which is precisely what drives the accelerated expansion of the universe. In order to satisfy the strong energy condition (SEC) we require the NEC to be satisfied, and in addition $\rho + p_\parallel + 2p_\perp = 2p_\perp \ge 0$. To satisfy strong energy condition (SEC), we add one more inequality to the null energy condition (NEC):
\begin{equation}
2p_\perp = \frac{e^{2\alpha(r)}}{4\pi}\Big(2\dot{\alpha}(r)^2 + \ddot{\alpha}(r) + \frac{2\dot{\alpha}(r)}{r}\Big) \ge 0.
\label{SET_4} 
\end{equation}

\textbf{The Dominant Energy Condition (DEC)} adds the requirement that the energy flux is non-spacelike (i.e., it does not travel faster than light). It ensures that energy does not flow in a causally prohibited way and implies that the WEC holds. For dominant energy condution (DEC) $\rho \geq |p_{ii}|$ for all $i = 1, 2, 3$.

The stress-energy tensor is the fundamental language in which matter and energy speak to spacetime, dictating its curvature and dynamics. The classical energy conditions were our first attempt to impose a grammar of "reasonableness" on this language, ensuring that the resulting narrative of the universe was causally sound and physically intuitive. The discovery that quantum mechanics systematically violates these conditions has not rendered them obsolete, but has instead deepened their significance. They now serve as a critical boundary, demarcating the well-trodden territory of classical physics from the wild and speculative frontiers of quantum gravity and cosmology. The ongoing dialogue between the precise formalism of $T_{\mu \nu}$ and the evolving principles of the energy conditions continues to guide our exploration of the universe's most profound secrets, from the singularities hidden within black holes to the possibility of spacetime itself being engineered.

\section{$\alpha(r) = -\Big(\frac{M}{r + l}\Big)$}

In this case the metric (\ref{6A_44}) is given by
\begin{eqnarray}
ds^2 = -e^{-\frac{2M}{r + l}}dt^2 + e^{\frac{2M}{r + l}}dr^2 + r^2\Big(d\theta^2 + sin^2\theta d\varphi^2\Big),
 \label{1A_55} 
\end{eqnarray}
and the metric component $g_{tt} = e^{-\frac{2M}{r + l}}$ is asymptoticaly flat:
\begin{eqnarray}
e^{-\frac{2M}{r+l}}\Big|_{r \to +\infty}  = 1
 \label{2A_55} 
\end{eqnarray}
and has following asymptotic behavior like a Schwarzschild metric component for big enough r:
\begin{eqnarray}
e^{-\frac{2M}{r+l}} \sim \Bigg(1 - \frac{2M}{r}\Bigg) + \mathcal{O}\Big(\frac{1}{r^2}\Big).
 \label{2A_55A} 
\end{eqnarray}

The peculiarity of this metric is, firstly, as can be seen from (\ref{2A_55A}), it closely resembles the Schwarzschild metric at big enough $r$. Secondly, in the metric component, the radial coordinate $r$ is raised to an exponential power, so this metric does not have any event horizon in the radial interval $r \in (0, +\infty)$ (see Fig.1).

\begin{figure}[h]
\center{\includegraphics[scale=0.2]{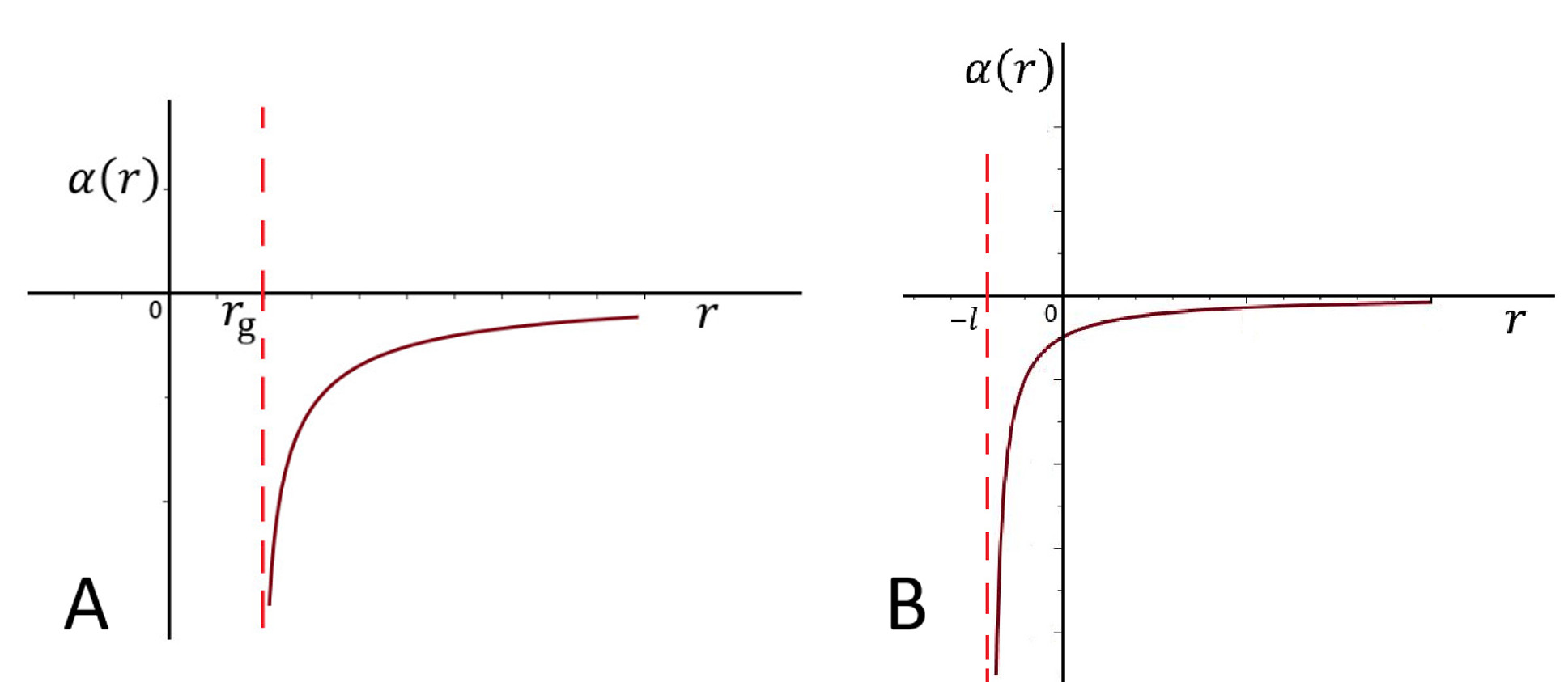}}
\caption{\textsl{The behavior of the function $\alpha(r)$ in the Schwarzschild ($\bf{A}$) and asymptotically Schwarzschild-like ($\bf{B}$) metrics. In the Schwarzschild metric, the function $\alpha(r)$  tends to $-\infty$ as $r \to r_g$, meaning there is an event horizon within the interval $r \in (0, +\infty)$. In the asymptotically Schwarzschild-like metric the function $\alpha(r)$   tends to $-\infty$ as $r \to -l$, meaning the event horizon lies outside the interval $r \in (0, +\infty)$, and this metric has no coordinate singularity within this interval.}} 
\label{Fig_1}
\end{figure}

Let us now find the photon sphere for massless particles and the ISCO for massive particles in the metric (\ref{1A_55}). For the photon orbit from (\ref{6F_44}) we obtain that
\begin{equation}
%\begin{gathered}
 \dot{V_0}\left(r\right) =\frac{L^2\Big(Mr - (r+l)^2\Big)}{(r+l)^2r^3}e^{-\frac{2M}{r+l}}.       \label{3A_55} 
%\end{gathered}
 \end{equation}

By solving equation $ \dot{V_0}\left(r\right) = 0$, we get following result for the photon sphere orbit $r_{\ast}$:
\begin{eqnarray}
r_{\ast} = \frac{1}{2}\Big(M - 2l \pm \sqrt{M^2 - 4Ml}\Big),
 \label{4A_55} 
\end{eqnarray}
 and as it is seen from this formula, the real value of $r_{\ast}$ is defined only in the interval of $l \in [0 , \frac{M}{4}]$.
For $l =0$, the roots are:
\begin{eqnarray}
r_{\ast1} = M \quad \text{or} \quad r_{\ast2} = 0, \nonumber 
\end{eqnarray}
and we choose $r_{\ast1} = M$ (the nonzero one). So the physical photon sphere radius is:
\begin{eqnarray}
r_{\ast} = \frac{1}{2}\Big(M - 2l + \sqrt{M^2 - 4Ml}\Big)
 \label{4B_55} 
\end{eqnarray}

and to verify stability, we must check the sign of $\ddot{V}_0(r)$.

The innermost stable circular orbit (ISCO) is a key concept in astrophysics, defining the boundary between stable and unstable orbits around compact objects. Within the ISCO, gravitational forces prevent stable motion, causing any particle to rapidly fall inward. This radius effectively sets the accretion disk's inner edge, thereby governing its structure and emission properties. To avoid a complex solution for $r$ , we instead assume a fixed circular orbit at $r = r_c$  and solve for the required angular momentum $L_c(r_c, M, l)$. The innermost circular orbit is then found by minimizing $L_c$  with respect to $r_c$,  given $ \dot{V_1}\left(r\right) = 0$. By using (\ref{7G89R}), we obtain
\begin{eqnarray}
\begin{gathered}
L_c^2 = \frac{\dot{\alpha}(r)r^3}{1 - \dot{\alpha}(r)r}, \label{5A_55} 
\end{gathered}
\end{eqnarray}
and solving equation $\frac{\partial L_c}{\partial r_c} =0$, we can find the innermost orbit $r_c$:
\begin{eqnarray}
r_{ISCO} = r_c =M - 2l \pm \sqrt{M^2 - 4Ml + l^2}.
 \label{6A_55} 
\end{eqnarray}
For $l =0$, the roots are:
\begin{eqnarray}
r_{c1} = 2M \quad \text{or} \quad r_{c2} = 0, \nonumber 
\end{eqnarray}
and we choose $r_{c1} = 2M$ (the nonzero one). So the physical ISCO is:
\begin{eqnarray}
r_{ISCO} = r_c =M - 2l  +  \sqrt{M^2 - 4Ml + l^2},
 \label{6A_55BB} 
\end{eqnarray}
 and as it is seen from this formula, the real value of $r_{ISCO}$ is defined only in the interval of $l \in [0 , \frac{M}{(\sqrt{3} + 2)}]$.

\subsection{Energy conditions}

Let us examine the Einstein field equations for this spacetime. The Stress-Energy Tensor ($T_\mu^\nu$) is the fundamental description of matter and energy in General Relativity, acting as the source of gravity. Energy Conditions are a set of rules used to define what "physically reasonable" matter is. They are essential tools for proving general theorems about the structure of spacetime, the existence of singularities, and the impossibility of certain time machines. Using $\rho = -T_t^t$, $p_{\parallel} = T_r^r$, $p_\perp = T_\theta^\theta = T_\varphi^\varphi$ and the mixed components $G_\mu^\nu = 8\pi T_\mu^\nu$, this yields the following form of the stress-energy-momentum tensor:
\begin{eqnarray}
\rho =\frac{1}{8\pi}\cdot \frac{1 - \Big(1 + 2\dot{\alpha}(r)\Big)e^{2\alpha(r)}}{r^2} =\frac{1}{8\pi} \cdot\frac{1 - \Big(1 + \frac{2Mr}{(r + l)^2}\Big)e^{-\frac{2M}{r + l}}}{r^2},    \nonumber  \\
p_\parallel =  \frac{1}{8\pi}\cdot\frac{\Big(1 + 2\dot{\alpha}(r)\Big)e^{2\alpha(r)} - 1}{r^2} =  \frac{1}{8\pi}\cdot\frac{\Big(1 + \frac{2Mr}{(r + l)^2}\Big)e^{-\frac{2M}{r + l}} - 1}{r^2}, \nonumber \\
p_\perp = \frac{1}{8\pi}\cdot e^{2\alpha(r)}\Big(2\dot{\alpha}(r)^2 + \ddot{\alpha}(r) + \frac{2\dot{\alpha}(r)}{r}\Big) = \frac{1}{8\pi}\cdot\frac{2(rM + rl +l)e^{-\frac{2M}{r + l}}M}{r(r + l)^4}.     
\label{EC_5_7} 
\end{eqnarray}

A necessary and suffcient conditions for the null energy condition (NEC) are: $\rho + p_\parallel \ge 0$ and $\rho + p_\perp \ge 0$. Since $\rho = -p_\parallel$, the first inequality is trivially satisfied. For the second condition we obtain from (\ref{EC_5_7}) following inequality:
\begin{eqnarray}
\rho + p_\perp  =\frac{1}{8\pi}\cdot \Bigg(\frac{1 - \Big(1 + \frac{2Mr}{(r + l)^2}\Big)e^{-\frac{2M}{r + l}}}{r^2} + \frac{2(rM + rl +l)e^{-\frac{2M}{r + l}}M}{r(r + l)^4}\Bigg) \geq 0.   
\label{EC_5_8} 
\end{eqnarray}
For big enough value of $u$ we get asymptotic relation:
\begin{eqnarray}
\rho + p_\perp  = \frac{4M^2}{8\pi r^4}\Big(1 + \frac{l}{M}\Big) + \mathcal{O}\Big(\frac{1}{r^5}\Big)  > 0,  
\label{EC_5_9} 
\end{eqnarray}
which reported that the second condition of the NEC is satisfied.

On the other hand, when $r$ is small, we obtain
\begin{eqnarray}
\rho + p_\perp  \sim  \frac{1 - e^{-\frac{2M}{l}}}{r^2}  > 0.
\label{EC_5_10} 
\end{eqnarray}

Therefore, for big enough and small values of $r$ the NEC is completely satisfied. If in addition the condition (\ref{SET_3}) is satisfied, we obtain satisfied Weak Energy condition (WEC) at the big enough and small values of $r$. But $\dot{\alpha}(r)=\frac{2M}{(r+l)^2}e^{-\frac{2M}{r + l}} > 0$, so everywhere $\rho < 0$, which dictated that the WEC is violated.

The Strong Energy Condition (SEC) is a hypothesis about the type of matter and energy that are "common" or "reasonable" in the Universe. It essentially states that gravity should always be attractive, and that the energy density measured by any observer must be non-negative and also dominate any pressure (or stress) in the system. In this case the additional condition for the SEC is (seen from (\ref{SET_4})):
\begin{equation}
2p_\perp = \frac{e^{2\alpha(r)}}{4\pi}\Big(2\dot{\alpha}(r)^2 + \ddot{\alpha}(r) + \frac{2\dot{\alpha}(r)}{r}\Big) = \frac{1}{8\pi}\cdot\frac{2(rM + rl +l)e^{-\frac{4M}{r + l}}M}{r(r + l)^4}   \ge 0. 
\label{EC_5_11} 
\end{equation}
It is easy to see that the expression in formula (\ref{EC_5_11}) always has a positive value in the interval $r \in (0, +\infty)$. So, the Strong Energy Condition (SEC) is satisfied at the big enough and small values of $r$.

The Dominant Energy Condition (DEC) is a rule we impose on the type of "stuff" that we believe should be allowed in our universe, according to our best understanding of general relativity. In essence, the DEC states that: Energy should not flow faster than the speed of light, and the energy density should be positive from any observer's point of view. The DEC is defined as:
\begin{eqnarray}
\rho - |p_\parallel| \geq 0, \qquad \rho - |p_\perp| \geq 0.
\label{EC_5_12} 
\end{eqnarray}

We see that $\rho - |p_\parallel| = 0$ and get following asymptotic relation for big enough $r$:
\begin{eqnarray}
\rho - |p_\perp| \sim \frac{2M\Big(l^2 + 2Ml + \frac{2M^2}{3}\Big)}{r^5} + \mathcal{O}\Big(\frac{1}{r^8}\Big).
\label{EC_5_13} 
\end{eqnarray}

On the other hand, when $r$ is small, we obtain
\begin{eqnarray}
\rho -  p_\perp  \sim  \frac{1 - e^{-\frac{2M}{l}}}{r^2}  > 0.
\label{EC_5_14} 
\end{eqnarray}
Therefore, it can be concluded that the DEC is satisfied at the big enough and small values of $r$.

\subsection{Regge–Wheeler analysis}

The Regge-Wheeler equation is a foundational result in the field of black hole perturbation theory. It represents the first successful attempt to describe the dynamics of a black hole when it is slightly disturbed. In essence, the Regge-Wheeler equation is a Schrödinger-type wave equation that describes how odd-parity (axial) gravitational waves scatter off a black hole. The Regge-Wheeler equation's importance cannot be overstated. It was the key that unlocked the linear dynamics of black holes and paved the way for:

- The understanding of gravitational wave signatures from perturbed black holes.

- The development of perturbation theory for rotating (Kerr) black holes, which is far more complex.

- Directly connecting theoretical predictions to observations made by gravitational wave detectors like LIGO and Virgo. When two black holes merge, the final black hole "rings down", and its signal is described by equations that are the direct descendants of the Regge-Wheeler equation.

The Regge–Wheeler equation for particles of spin $S \in \{0, 1\}$ is \cite{Boon}
\begin{eqnarray}
\frac{\partial^2\Psi}{\partial R_{\ast}^2}  + \Big \{ \omega^2 + \mathcal{V}_S \Big \}\Psi = 0,
 \label{RW1_1} 
\end{eqnarray}
where $R_{\ast}$ is tortoise coordinate:
\begin{eqnarray}
dR_{\ast} = e^{\frac{2M}{r +l}}dr \nonumber \\
\end{eqnarray}
which, gives the following expression for the metric (\ref{1A_55}):
\begin{eqnarray}
ds^2 = e^{-\frac{2M}{r + l}}\Bigg \{-dt^2 + dR_{\ast}^2 \Bigg \} + r^2\Big(d\theta^2 + \text{sin}^2\theta d\varphi^2\Big).
 \label{RW1_1A} 
\end{eqnarray}
It is useful to express this as
\begin{eqnarray}
ds^2 = A(R_{\ast})^2\Bigg \{-dt^2 + dR_{\ast}^2 \Bigg \} +  B(R_{\ast})^2\Big(d\theta^2 + \text{sin}^2\theta d\varphi^2\Big).
 \label{RW1_1B} 
\end{eqnarray}

The general formula for the Regge–Wheeler potential is \cite{Simp}:
\begin{eqnarray}
\mathcal{V}_S = \Bigg \{ \frac{A^2}{B^2}\Bigg \}\tilde{l}(\tilde{l} + 1) + (1-S)\frac{1}{B}\frac{\partial^2B}{\partial r_{\ast}^2}.
 \label{RW1_2} 
\end{eqnarray}

Note that for our metric $\frac{\partial}{\partial R_{\ast}} = e^{-\frac{2M}{r + l}}\frac{\partial}{\partial r} $ and $B = r$, we have
\begin{eqnarray}
\frac{1}{B}\frac{\partial^2B}{\partial R_{\ast}^2} = \frac{2M}{r(r+l)^2} e^{-\frac{4M}{r + l}}.
 \label{RW1_3} 
\end{eqnarray}
Therefore Regge–Wheeler potential has following form:
\begin{eqnarray}
\mathcal{V}_{S \in \{0, 1\}} =  e^{-\frac{2M}{r + l}}\Bigg[\frac{\tilde{l}(\tilde{l} + 1)}{r^2}  +(1 - S)\frac{2M}{r(r + l)^2}e^{-\frac{2M}{r + l}}\Bigg].
 \label{RW1_4} 
\end{eqnarray}

\subsection{Spin zero}

In particular for spin zero one 
\begin{eqnarray}
\mathcal{V}_{0} =  e^{-\frac{2M}{r + l}}\Bigg[\frac{\tilde{l}(\tilde{l} + 1)}{r^2}  +\frac{2M}{r(r + l)^2}e^{-\frac{2M}{r + l}}\Bigg].
 \label{RW1_5} 
\end{eqnarray}
We now apply the results obtained above to the Schwarzschild metric and obtain
\begin{eqnarray}
\mathcal{V}_{0, Sch} =  \Bigg(\frac{1 - \frac{M}{2r}}{1+\frac{M}{2r}}\Bigg)^2\Bigg[\frac{\tilde{l}(\tilde{l} + 1)}{r^2\Big(1 + \frac{M}{2r}\Big)^4}  +\frac{2M}{r^3(1 + \frac{M}{2r})^6}\Bigg].
 \label{RW1_6} 
\end{eqnarray}
The s-wave ($\tilde{l} = 0$) is the most physically important case for scalar fields (see Fig.2)
\begin{eqnarray}
\mathcal{V}_{0, \tilde{l} = 0} =  \frac{2M}{r(r + l)^2}e^{-\frac{4M}{r + l}}
 \label{RW1_7} 
\end{eqnarray}
and 
\begin{eqnarray}
\mathcal{V}_{0, \tilde{l} = 0, Sch} =  \Bigg(\frac{1 - \frac{M}{2r}}{1+\frac{M}{2r}}\Bigg)^2\Bigg[\frac{2M}{r^3(1 + \frac{M}{2r})^6}\Bigg].
 \label{RW1_8} 
\end{eqnarray}

The function (\ref{RW1_7}) never vanishes and has two extreme points in the interval $r \in (0, +\infty)$:
\begin{eqnarray}
r_{1,2} = \frac{2(M - l) \pm \sqrt{l^2 - 8Ml + 4M^2}}{3}.
 \label{RW1_8NN} 
\end{eqnarray}
The potential peak and local minimum are at $r_{1} = \frac{2(M - l) + \sqrt{l^2 - 8Ml + 4M^2}}{3}$ and $r_{2} = \frac{2(M - l) - \sqrt{l^2 - 8Ml + 4M^2}}{3}$  respectively. The potential peak point for the Schwarzschild metric is $r_3 = \frac{3M}{2}$.  In case of $l = M(4 - 2\sqrt{3})$, the function (\ref{RW1_7}) has a single extreme point:
\begin{eqnarray}
r_{4} = \frac{2(M - l)}{3}.
 \label{RW1_8NN} 
\end{eqnarray}
If $l > M(4 - 2\sqrt{3})$, the function has no any extreme point in $r \in (0, +\infty)$. 

\begin{figure}[h]
\center{\includegraphics[scale=0.2]{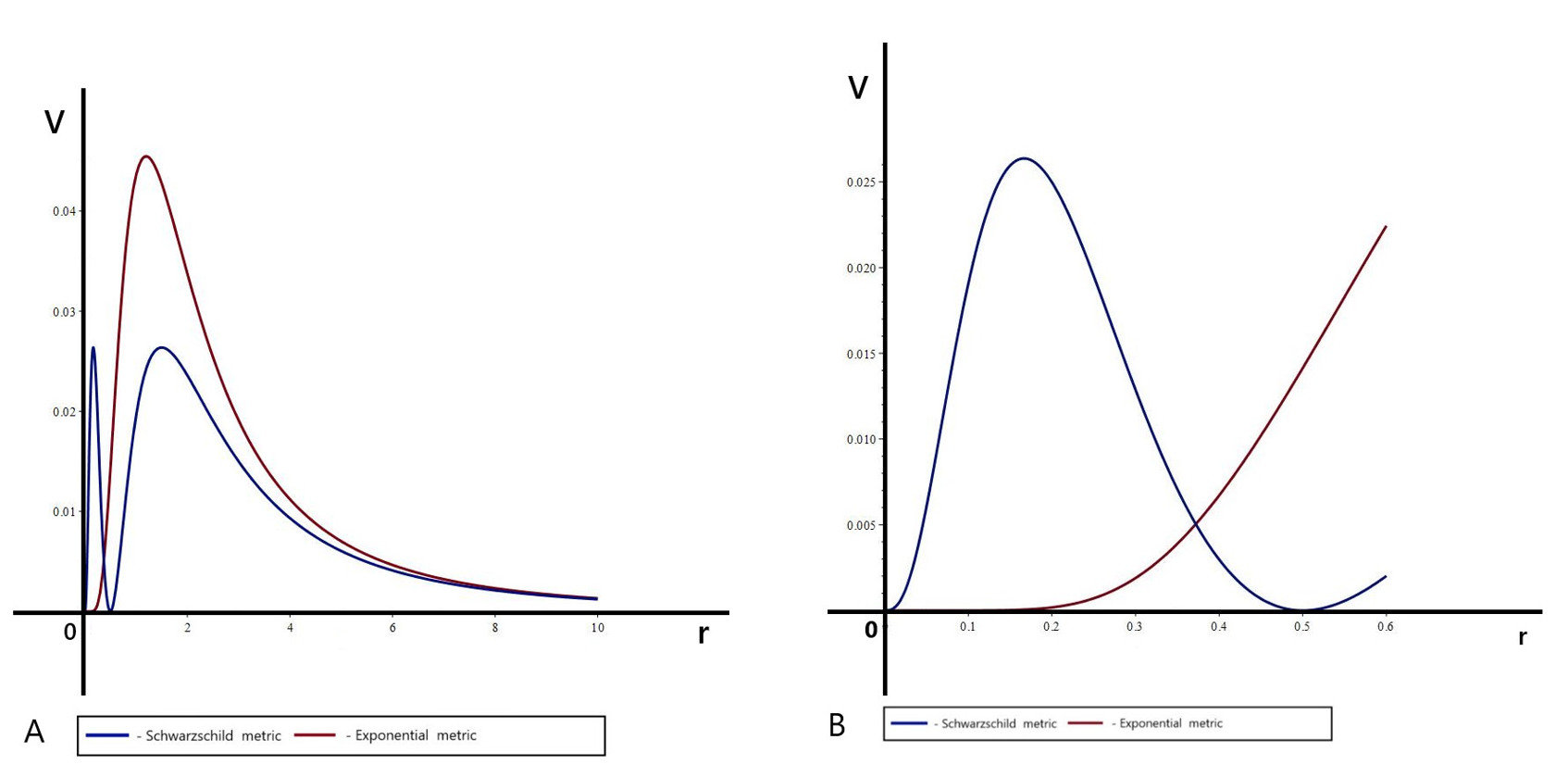}}
\caption{\textsl{The potential function for $M=1$ and $l=0.1$. The graph shows the spin-zero Regge–Wheeler potential for $\tilde{l} = 0$. While these potentials are similar at large $r$, they differ radically at small $r$. In the Schwarzschild case, however, the potential at small $r$ is only formal, as this region lies behind the event horizon and cannot interact with the domain of outer communication.}} 
\label{Fig_2}
\end{figure}

\subsection{Spin one}

For the spin one case  (\ref{RW1_4}) is converted into
\begin{eqnarray}
\mathcal{V}_{1} =  e^{-\frac{2M}{r + l}}\Bigg[\frac{\tilde{l}(\tilde{l} + 1)}{r^2}  \Bigg],
 \label{RW1_9} 
\end{eqnarray}
and in the Schwarzschild metric (see Fig.3), the result is
\begin{eqnarray}
\mathcal{V}_{1, Sch} =  \Bigg(\frac{1 - \frac{M}{2r}}{1+\frac{M}{2r}}\Bigg)^2\Bigg[\frac{\tilde{l}(\tilde{l} + 1)}{r^2\Big(1 + \frac{M}{2r}\Big)^4} \Bigg].
 \label{RW1_10} 
\end{eqnarray}

\begin{figure}[h]
\center{\includegraphics[scale=0.2]{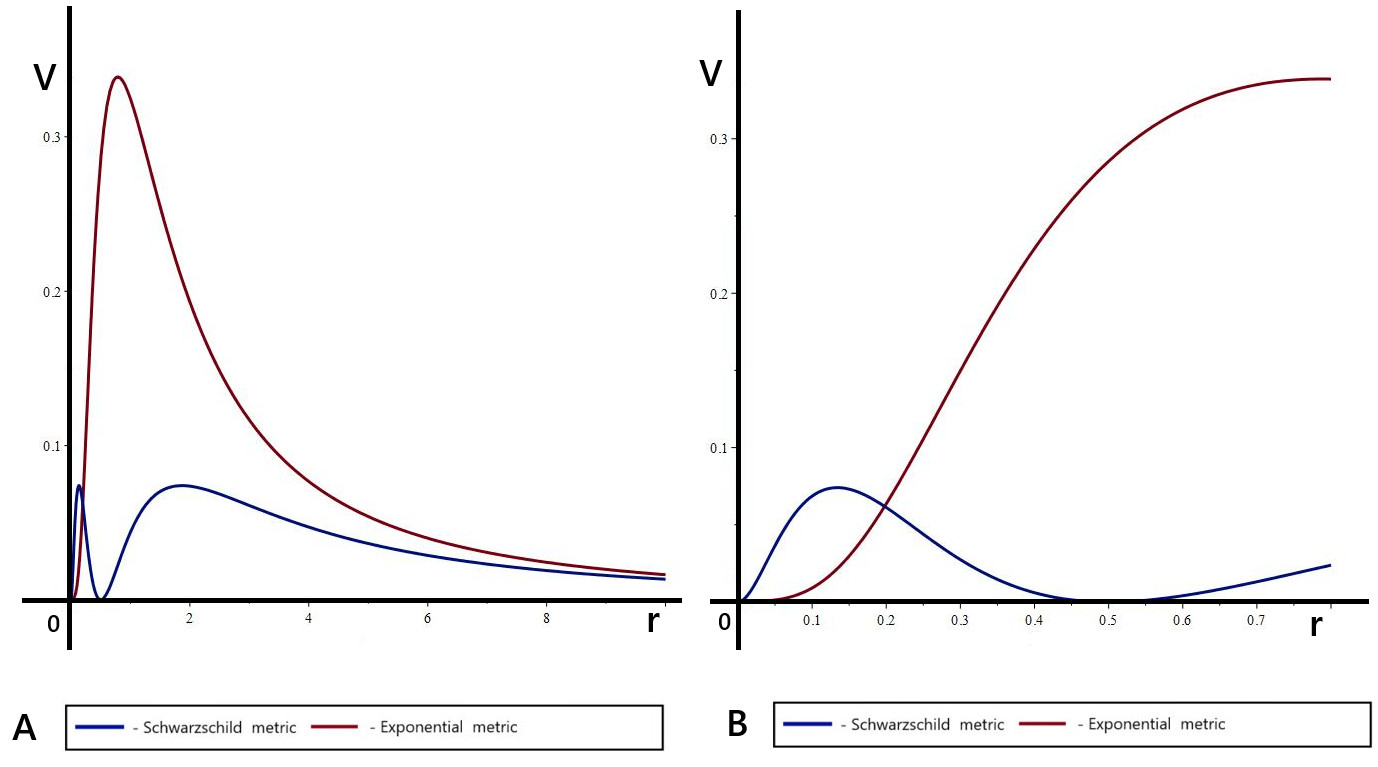}}
\caption{\textsl{The potential function for $M =1$ and $l = 0.1$. The graph shows the spin one Regge–Wheeler potential for $\tilde{l} = 1$. While these potentials are similar at large $r$, they differ radically at small $r$. In the Schwarzschild case, however, the potential at small $r$ is only formal, as this region lies behind the event horizon and cannot interact with the domain of outer communication.}} 
\label{Fig_3}
\end{figure}

\section{$\alpha(r) = \frac{M}{l}\Big[arctan\Big(\frac{r}{l}\Big) - \frac{\pi}{2}\Big]$}

In this case the metric (\ref{6A_44}) is given by
\begin{eqnarray}
ds^2 = -e^{\frac{2M}{l}\Big[arctan\Big(\frac{r}{l}\Big) - \frac{\pi}{2}\Big]}dt^2 + e^{-\frac{2M}{l}\Big(arctan\Big(\frac{r}{l}\Big) - \frac{\pi}{2}\Big)}dr^2 + r^2\Big(d\theta^2 + \text{sin}^2\theta d\varphi^2\Big),
 \label{1A_666} 
\end{eqnarray}
and the metric component $g_{tt} = e^{-\frac{2M}{r + l}}$ has following asymptotic behavior for big enough r:
\begin{eqnarray}
e^{\frac{2M}{l}\Big[arctan\Big(\frac{r}{l}\Big) - \frac{\pi}{2}\Big]} \sim \Bigg(1 - \frac{2M}{r}\Bigg) + \mathcal{O}\Big(\frac{1}{r^2}\Big).
 \label{2A_66A} 
\end{eqnarray}
The peculiarity of this metric is, firstly, as can be seen from (\ref{2A_55}), it closely resembles the Schwarzschild metric at big enough $r$. Secondly, in the metric component, the radial coordinate $r$ is raised to an exponential power, so this metric does not have any event horizon in the radial interval $r \in (0, +\infty)$ (see Fig.4).

\begin{figure}[h]
\center{\includegraphics[scale=0.2]{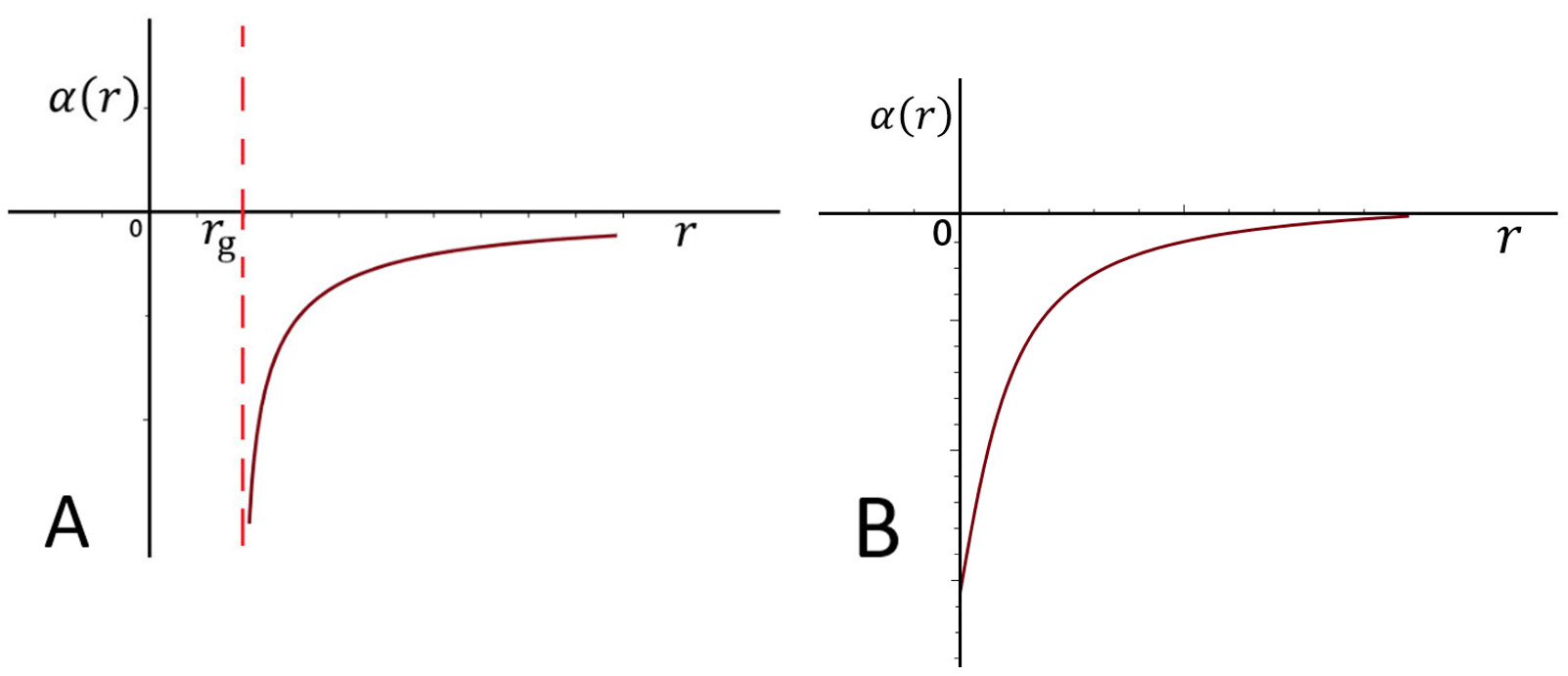}}
\caption{\textsl{The behavior of the function $\alpha(r)$ in the Schwarzschild ($\bf{A}$) and asymptotically Schwarzschild-like ($\bf{B}$) metrics. In the Schwarzschild metric, the function $\alpha(r)$  tends to $-\infty$ as $r \to r_g$, meaning there is an event horizon within the interval $r \in (0, +\infty)$. The asymptotically Schwarzschild-like metric has no any event horizon in the interval $r \in (0, +\infty)$.}} 
\label{Fig_4}
\end{figure}

Let us now find the photon sphere for massless particles and the ISCO for massive particles in the metric (\ref{1A_55}). For the photon orbit from (\ref{6F_44}) we obtain that
\begin{equation}
%\begin{gathered}
 \dot{V_0}\left(r\right) =\frac{L^2\Big(Mr - l^2 - r^2\Big)}{(r+l)^2r^3}e^{-\frac{2M}{l}\Big[arctan\Big(\frac{r}{l}\Big) - \frac{\pi}{2}\Big]}.     \label{3A_66A} 
%\end{gathered}
 \end{equation}

By solving equation $ \dot{V_0}\left(r\right) = 0$, we get following result for the photon sphere orbit $r_{\ast}$:
\begin{eqnarray}
r_{\ast} = \frac{M}{2}\Bigg(1  \pm \sqrt{1 - \Bigg(\frac{2l}{M}\Bigg)^2}\Bigg).
 \label{4A_66A} 
\end{eqnarray}
For $l =0$, the roots are:
\begin{eqnarray}
r_{\ast1} = M \quad \text{or} \quad r_{\ast2} = 0, \nonumber 
\end{eqnarray}
and we choose $r_{\ast1} = M$ (the nonzero one). So the physical ISCO is:
\begin{eqnarray}
r_{\ast} = \frac{M}{2}\Bigg(1  + \sqrt{1 - \Bigg(\frac{2l}{M}\Bigg)^2}\Bigg),
 \label{4A_66BBB} 
\end{eqnarray}
 and as it is seen from this formula, the real value of $r_{\ast}$ is defined only in the interval of $l \in [0 , \frac{M}{2}]$.

By using (\ref{7G89R}), we obtain
\begin{eqnarray}
\begin{gathered}
L_c^2 = \frac{Mr_c^3}{r_c^2 - Mr_c + l^2}, \label{5A_66A} 
\end{gathered}
\end{eqnarray}
and solving equation $\frac{\partial L_c}{\partial r_c} =0$, we can find the innermost orbit $r_c$:
\begin{eqnarray}
r_{ISCO} = r_c =M\Bigg(1 \pm \sqrt{1 -3\Bigg(\frac{l}{M}\Bigg)^2}\Bigg).
 \label{6A_66A} 
\end{eqnarray}
For $l =0$, the roots are:
\begin{eqnarray}
r_{c1} = 2M \quad \text{or} \quad r_{c2} = 0, \nonumber 
\end{eqnarray}
and we choose $r_{c1} = 2M$ (the nonzero one). So the physical ISCO is:
\begin{eqnarray}
r_{ISCO} = r_c =M\Bigg(1  + \sqrt{1 -3\Bigg(\frac{l}{M}\Bigg)^2}\Bigg),
 \label{6A_66ABBBB} 
\end{eqnarray}
 and as it is seen from this formula, the real value of $r_{ISCO}$ is defined only in the interval of $l \in [0 , \frac{M}{\sqrt{3}}]$.

\subsection{Energy conditions}

The Stress-Energy Tensor ($T_\mu^\nu$) is the fundamental description of matter and energy in General Relativity, acting as the source of gravity. Energy Conditions are a set of rules used to define what "physically reasonable" matter is. They are essential tools for proving general theorems about the structure of spacetime, the existence of singularities, and the impossibility of certain time machines. Let us examine the Einstein field equations for this spacetime. Using $\rho = -T_t^t$, $p_{\parallel} = T_r^r$, $p_\perp = T_\theta^\theta = T_\varphi^\varphi$ and the mixed components $G_\mu^\nu = 8\pi T_\mu^\nu$, this yields the following form of the stress-energy-momentum tensor:
\begin{eqnarray}
\rho = \frac{1}{8\pi}\cdot \frac{1 - \Big(1 + 2\dot{\alpha}(r)\Big)e^{2\alpha(r)}}{r^2} = \frac{1}{8\pi}\cdot \Bigg \{\Bigg(\frac{2M}{r(r^2 + l^2)} + \frac{1}{r^2}\Bigg)e^{\frac{2M}{l}\Big[arctan\Big(\frac{r}{l}\Big) - \frac{\pi}{2}\Big]} - \frac{1}{r^2}\Bigg \},     \nonumber  \\
p_\parallel = \frac{1}{8\pi}\cdot\frac{\Big(1 + 2\dot{\alpha}(r)\Big)e^{2\alpha(r)} - 1}{r^2} = \frac{1}{8\pi}\cdot \Bigg \{\Bigg( \frac{1}{r^2} - \frac{2M}{r(r^2 + l^2)} + \frac{1}{r^2}\Bigg)e^{\frac{2M}{l}\Big[arctan\Big(\frac{r}{l}\Big) - \frac{\pi}{2}\Big]}\Bigg \},       \nonumber \\
p_\perp = \frac{e^{2\alpha(r)}}{8\pi} \cdot\Big(2\dot{\alpha}(r)^2 + \ddot{\alpha}(r) + \frac{2\dot{\alpha}(r)}{r}\Big) = \frac{1}{8\pi}\cdot \frac{2M(Mr + l^2)e^{\frac{2M}{l}\Big[arctan\Big(\frac{r}{l}\Big) - \frac{\pi}{2}\Big]}}{r(r^2 + l^2)^2}.  \qquad   
\label{EC_6_7A66} 
\end{eqnarray}

A necessary and sufficient set of conditions for the null energy condition (NEC) is given by $\rho + p_\parallel \ge 0$ and $\rho + p_\perp \ge 0$. The equality $\rho = -p_\parallel$ ensures the first condition is identically true. The second condition, derived from Eq. (\ref{EC_6_7A66}), leads to the inequality:
\begin{eqnarray}
\rho + p_{\perp} = \frac{1}{8\pi}\cdot \Bigg[\Bigg \{\Bigg(\frac{2M}{r(r^2 + l^2)} + \frac{1}{r^2}\Bigg)e^{\frac{2M}{l}\Big[arctan\Big(\frac{r}{l}\Big) - \frac{\pi}{2}\Big]} - \frac{1}{r^2}\Bigg \}  +\\ \nonumber
+ \frac{2M(Mr + l^2)e^{\frac{2M}{l}\Big[arctan\Big(\frac{r}{l}\Big) - \frac{\pi}{2}\Big]}}{r(r^2 + l^2)^2}\Bigg] \geq 0. \quad   
\label{EC_6_8A6} 
\end{eqnarray}

For big enough value of $r$ we get asymptotic relation:
\begin{eqnarray}
\rho + p_\perp  = \Big(\frac{2M}{r^2}\Big)^2 + \mathcal{O}\Big(\frac{1}{r^5}\Big)  > 0,  
\label{EC_6_9A6} 
\end{eqnarray}
which reported that the second condition of the NEC is satisfied.

On the other hand, when $r$ is small, we obtain
\begin{eqnarray}
\rho + p_\perp  \sim  \frac{e^{-\frac{2M}{l}} - 1}{r^2} +  \frac{4Me^{-\frac{M\pi}{l}}}{l^2r} + \mathcal{O}\Big(1\Big).
\label{EC_6_10A6} 
\end{eqnarray}
When
\begin{eqnarray}
r > r_0 = \frac{l^2\Big(1 - e^{\frac{2M}{l}} \Big)}{4Me^{-\frac{M\pi}{l}}}  \nonumber 
\end{eqnarray}
expression in (\ref{EC_6_10A6}) has a positive value.

We can therefore conclude that the Null Energy Condition (NEC) holds for both the large and small values of $r$ untill $r_0$. With the additional requirement that condition (\ref{SET_3}) is satisfied, the Weak Energy Condition (WEC) is also fulfilled in these regimes. As our cumbersome computations show, condition (\ref{SET_3}) is satisfied only for small values of r. It is reported that the WEC is satisfied only in small values of $r$ till $r_0$.

As it is shown in our calculations, the additional condution (\ref{SET_4}) is always satisfied:
\begin{equation}
2p_\perp = \frac{e^{2\alpha(r)}}{4\pi}\Big(2\dot{\alpha}(r)^2 + \ddot{\alpha}(r) + \frac{2\dot{\alpha}(r)}{r}\Big) =  \frac{1}{8\pi}\cdot \frac{4M(Mr + l^2)e^{\frac{2M}{l}\Big[arctan\Big(\frac{r}{l}\Big) - \frac{\pi}{2}\Big]}}{r(r^2 + l^2)^2} > 0      
\label{EC_6_11A6} 
\end{equation}
in the interval of $r \in (0, +\infty)$.

From there, it can be concluded that the Strong Energy Condition (SEC) is satisfied for both the large and small values of $r$ untill $r_0$. 

The DEC is defined as:
\begin{eqnarray}
\rho - |p_\parallel| \geq 0, \qquad \rho - |p_\perp| \geq 0. \nonumber
\end{eqnarray}

We see that $\rho - |p_\parallel| = 0$ and get following asymptotic relation for big enough $r$:
\begin{eqnarray}
\rho - |p_\perp| \sim - \Bigg(\frac{2M}{r^2}\Bigg)^2 + \mathcal{O}\Bigg(\frac{1}{r^5}\Bigg) < 0.
\label{EC_6_13A6} 
\end{eqnarray}

On the other hand, when $r$ is small, we obtain
\begin{eqnarray}
\rho - |p_\perp|  \sim  \frac{e^{-\frac{2M}{l}} - 1}{r^2} +  \frac{2Me^{-\frac{M\pi}{l}}}{l^2r} + \mathcal{O}\Big(1\Big) <0
\label{EC_6_10F6} 
\end{eqnarray}
for small value of $r$. This means that the DEC is violated for big enough and small values of $r$.

\subsection{Regge–Wheeler analysis}

A tortoise coordinate is defined as follows:
\begin{eqnarray}
dR_{\ast} = e^{\frac{2M}{l}\Big[arctan\Big(\frac{r}{l}\Big) - \frac{\pi}{2}\Big]}dr, \nonumber \\
\end{eqnarray}
which, gives the following expression for the metric (\ref{1A_666}):
\begin{eqnarray}
ds^2 = e^{\frac{2M}{l}\Big[arctan\Big(\frac{r}{l}\Big) - \frac{\pi}{2}\Big]}\Bigg \{-dt^2 + dR_{\ast}^2 \Bigg \} + r^2\Big(d\theta^2 + \text{sin}^2\theta d\varphi^2\Big).
 \label{RW2_1B} 
\end{eqnarray}
It is useful to express this as
\begin{eqnarray}
ds^2 = A(R_{\ast})^2\Bigg \{-dt^2 + dR_{\ast}^2 \Bigg \} +  B(R_{\ast})^2\Big(d\theta^2 + \text{sin}^2\theta d\varphi^2\Big).
 \label{RW2_2B} 
\end{eqnarray}

Using the general formula for the Regge-Wheeler potential (\ref{RW1_2}) and note that for this metric $\frac{\partial}{\partial R_{\ast}} = e^{\frac{2M}{l}\Big[arctan\Big(\frac{r}{l}\Big) - \frac{\pi}{2}\Big]}\frac{\partial}{\partial r} $ and $B = r$, we have
\begin{eqnarray}
\frac{1}{B}\frac{\partial^2B}{\partial R_{\ast}^2} = \frac{2M}{r(r^2 + l^2)}e^{\frac{2M}{l}\Big[arctan\Big(\frac{r}{l}\Big) - \frac{\pi}{2}\Big]}.
 \label{RW1_3} 
\end{eqnarray}
Therefore Regge–Wheeler potential has following form:
\begin{eqnarray}
\mathcal{V}_{S \in \{0, 1\}} =  e^{\frac{2M}{l}\Big[arctan\Big(\frac{r}{l}\Big) - \frac{\pi}{2}\Big]}\Bigg[\frac{\tilde{l}(\tilde{l} + 1)}{r^2}  +(1 - S)\frac{2M}{r(r^2 + l^2)}e^{\frac{2M}{l}\Big[arctan\Big(\frac{r}{l}\Big) - \frac{\pi}{2}\Big]}\Bigg].
 \label{RW2_Zero} 
\end{eqnarray}

\subsection{Spin zero}

We now apply the results from (\ref{RW2_Zero}) to the spin-zero case 
\begin{eqnarray}
\mathcal{V}_{0} =  e^{\frac{2M}{l}\Big[arctan\Big(\frac{r}{l}\Big) - \frac{\pi}{2}\Big]}\Bigg[\frac{\tilde{l}(\tilde{l} + 1)}{r^2}  +\frac{2M}{r(r^2 + l^2)}e^{\frac{2M}{l}\Big[arctan\Big(\frac{r}{l}\Big) - \frac{\pi}{2}\Big]}\Bigg].
 \label{RW2_Zero_A} 
\end{eqnarray}
The s-wave ($\tilde{l} = 0$) is the most physically important case for scalar fields (see Fig.5)
\begin{eqnarray}
\mathcal{V}_{0} =  e^{\frac{4M}{l}\Big[arctan\Big(\frac{r}{l}\Big) - \frac{\pi}{2}\Big]}\Bigg[\frac{2M}{r(r^2 + l^2)}\Bigg].
 \label{RW2_Zero_B} 
\end{eqnarray}

\begin{figure}[h]
\center{\includegraphics[scale=0.2]{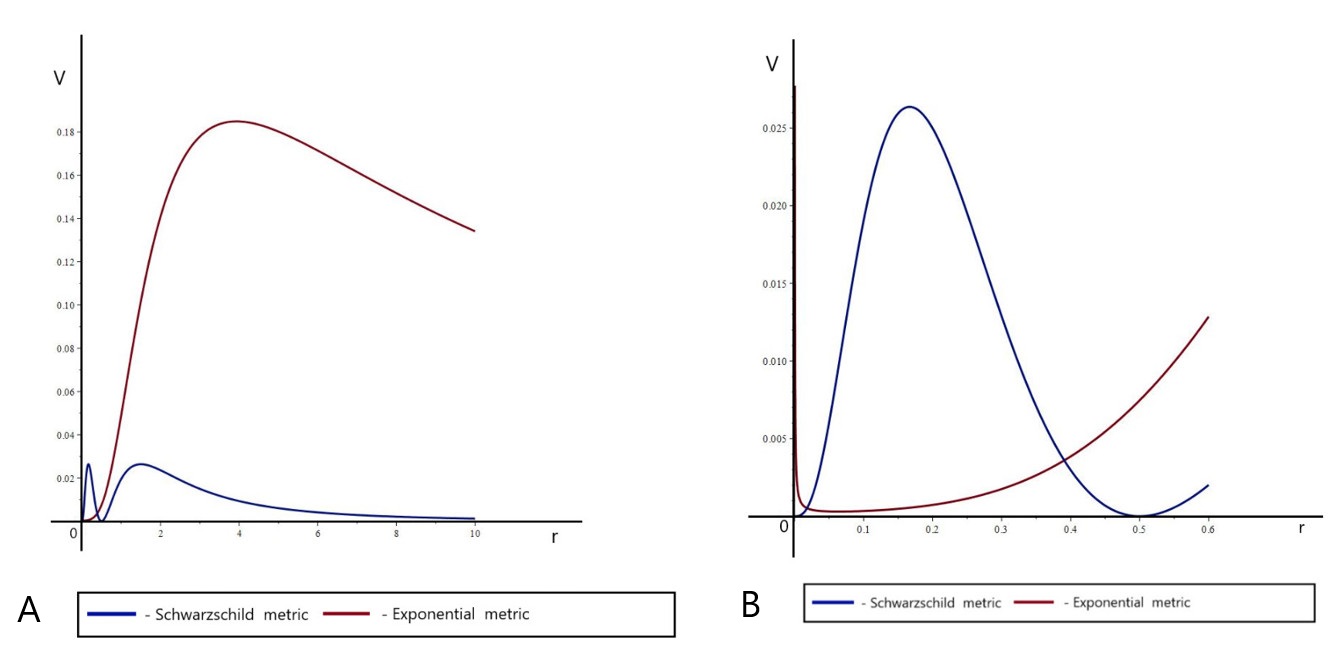}}
\caption{\textsl{The potential function for $M = 1$ and $l = 0.5$. The graph shows the spin zero Regge–Wheeler potential for $\tilde{l} = 0$. As it is shown in the graph they are different in all radial coordinate $r$. In the Schwarzschild case, however, the potential at small $r$ is only formal, as this region lies behind the event horizon and cannot interact with the domain of outer communication.}} 
\label{Fig_5}
\end{figure}

\subsection{Spin one}
For the spin one case  (\ref{RW2_Zero}) is converted into (see Fig.6)
\begin{eqnarray}
\mathcal{V}_{S \in \{0, 1\}} =  e^{\frac{2M}{l}\Big[arctan\Big(\frac{r}{l}\Big) - \frac{\pi}{2}\Big]}\Bigg[\frac{\tilde{l}(\tilde{l} + 1)}{r^2} \Bigg].
 \label{RW2_One} 
\end{eqnarray}

\begin{figure}[h]
\center{\includegraphics[scale=0.2]{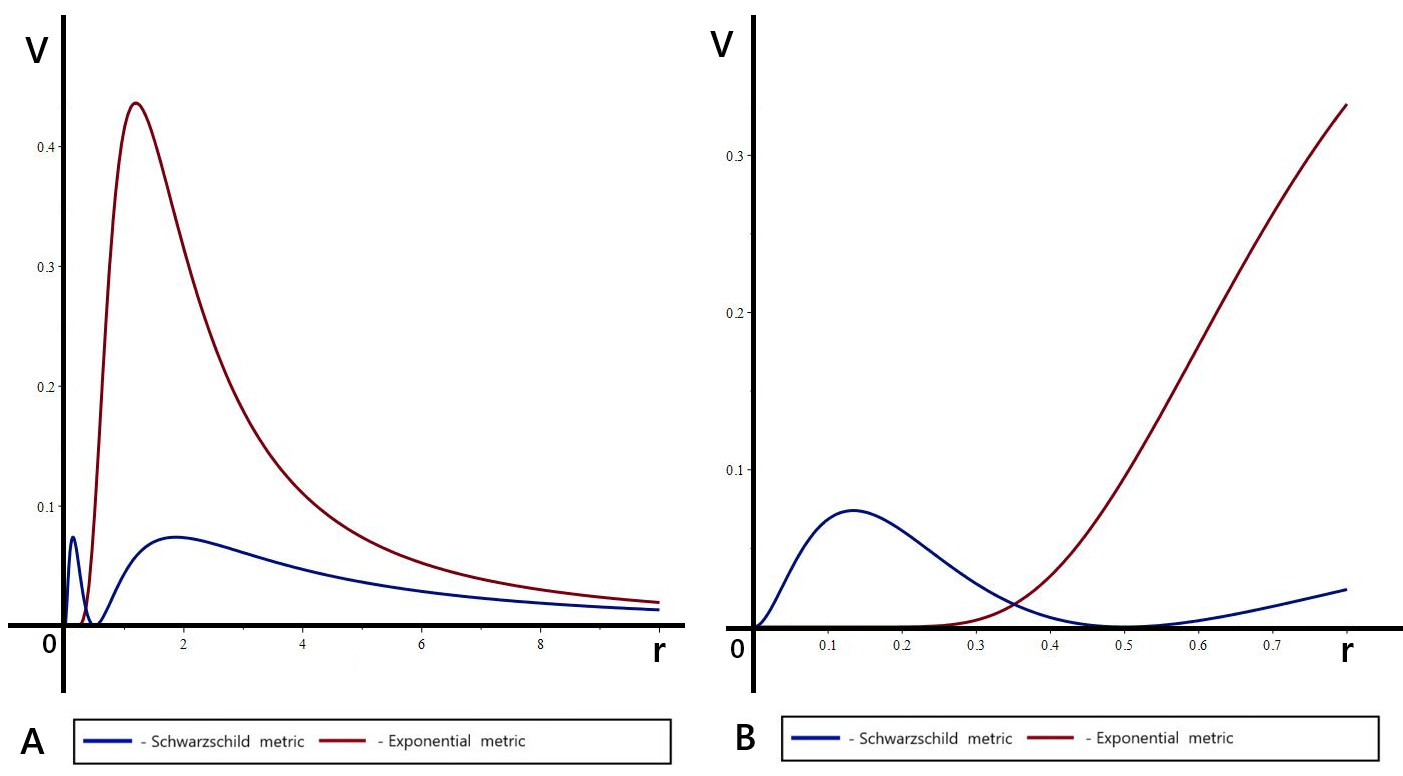}}
\caption{\textsl{The potential function for $M =1$ and $l = 1$. The graph shows the spin one Regge–Wheeler potential for $\tilde{l} = 1$. While these potentials are similar at large $r$, they differ radically at small $r$. In the Schwarzschild case, however, the potential at small $r$ is only formal, as this region lies behind the event horizon and cannot interact with the domain of outer communication. field.}} 
\label{Fig_6}
\end{figure}

\section{Geodesic equations}

With the metric (\ref{6A_44}) in mind, we now outline several useful relations for geodesics in a generic static, spherically-symmetric spacetime using the Buchdal (quasi-global) parametrization:
\begin{equation} \label{b4.m}
 ds^2 =- A(r) dt ^2 + \frac{dr^2}{A(r)} + C(r) d \Omega^2,
\end{equation}
where $A = A(r) > 0$ is the red-shift function, $C(r) > 0$ is the central function, 
$d\Omega^2=d\theta^2+\sin^2\theta d\phi^2$ is the metric of the $2$-dimensional sphere.

\subsection{Circular null geodesics}

Now let us consider the application of the results obtained in \cite{Ivash_2025_3} for these metrics. Consider null geodesics with non-zero angular momentum
$L \neq 0$. In this case the effective potential \eqref{8.F_44} 
has the following form
\begin{equation}\label{Unull}
U(r) = A(r)\frac{L^2}{r^2}.
\end{equation}
For the simplest case of circular orbits with a constant radius $r_\ast$
\begin{equation} \label{b.64}
r = r_* = \const, \quad L \neq 0,
\end{equation}
The trajectories form a sphere known as the photon sphere. From Eqs. \eqref{8F_66K}, \eqref{Unull}, and \eqref{b.64}, the radius $r_\ast$ of this sphere is determined by the relation 
\begin{equation}\label{b.65}
\frac{d}{dr}\left( \frac{A}{C} \right) \biggr|_{r = r_*} = 0.
\end{equation}
From Eqs. \eqref{7.AB} it follows that
\begin{equation} \label{b.67}
A(r_{\ast})\dot{t} = \varepsilon \rightarrow t - t_0 = \frac{\varepsilon}{A(r_*)}(\tau - \tau_0),
\end{equation}
\begin{equation}\label{b.68}
C(r_*)\dot{\phi} = L \rightarrow \phi - \phi_0 = \frac{L}{C(r_*)}(\tau - \tau_0),
\end{equation}
where $t_0,\tau_0,\phi_0$ are arbitrary constants. Thus,
\begin{equation}\label{b.69}
\phi - \phi_0 = \Omega(t - t_0),
\end{equation}
where
\begin{equation}\label{b.70}
\Omega = \frac{L A(r_*)}{\varepsilon C(r_*)}
\end{equation}
is the cyclic ``orbital'' frequency. Here and in what follows, we set for speed of light $c = 1$.

From Eq. \eqref{8.Y_4545} we obtain:
\begin{equation}\label{b.71}
U(r_{\ast}) = \varepsilon^2 \rightarrow \varepsilon = \pm |L|\sqrt{\frac{A(r_*)}{C(r_*)}},
\end{equation}
and hence
\begin{equation}
\label{b.72}
\Omega = \frac{lA(r_*)}{\varepsilon C(r_*)} =
\pm \frac{L}{|L|}\sqrt{\frac{A(r_*)}{C(r_*)}}.
\end{equation}

\subsection{Generic null geodesics}
We examine non-radial, non-circular trajectories of photons in the equatorial plane ($\theta = \pi/2$) that, for almost all values of the affine parameter (singular points excepted), satisfy:
\begin{equation}\label{b.81}
\dot r \neq 0,
\end{equation}
for which the radial equation \eqref{8F_55K} follows from \eqref{8.Y_4545}, also
expressible as:
\begin{equation}\label{b.82}
\frac{dr}{d\tau} = \pm \sqrt{\varepsilon^2 - U(r)}.
\end{equation}

Using Eqs. \eqref{7.AB}, we derive the relations
\begin{equation}\label{b.83}
\frac{dt}{dr} = \pm \frac{\varepsilon}
{A(r)\sqrt{\varepsilon^2 - U(r)}},
\end{equation}
\begin{equation}\label{b.84}
\frac{d\phi}{dr} = \pm \frac{L}{r^2\sqrt{\varepsilon^2 - U(r)}},
\end{equation}
whose integration with equation \eqref{b.82} leads to the formal quadratures for the geodesic equations when $(\dot r \neq 0)$:
\begin{equation}\label{b.85}
t - t_0 =  \pm \varepsilon\int_{r_0}^r\frac{d\bar r}{g_{tt}\left( \bar r \right)\sqrt{\varepsilon^2 - U(\bar r)}},
\end{equation}
\begin{equation}\label{b.86}
\phi - \phi_0 =  \pm L\int_{r_0}^r\frac{d\bar r}{C(\bar r)\sqrt{\varepsilon^2 - U(\bar r)}},
\end{equation}\label{b.87}
\begin{equation}
\theta = \frac{\pi}{2},
\end{equation}
\begin{equation}\label{b.88}
\pm \int_{r_0}^r\frac{d\bar r}{\sqrt{\varepsilon^2 - U(\bar r)}} = \tau - \tau_0.
\end{equation}
%Here, we recall that
%\begin{equation}\label{2.89}
%U(r) = A(r)\frac{l^2}{C(r)}.
%\end{equation}

\subsection{Spiral trajectory cases}

If we substitute into Eq. \eqref{b.86} the relation
\begin{equation}\label{b.90}
\varepsilon = \pm \sqrt{U(r_*)},
\end{equation}
which follows from \eqref{b.71} and corresponds to the circular motion, we obtain an infinite
spiral winding around the photon sphere with an infinite number of revolutions (a ``snail-like'' trajectory):
\begin{equation}\label{b.91}
\phi - \phi_0 =  \pm L\int_{r_0}^r\frac{d\bar r}{C(\bar r)\sqrt{U(r_*) - U(\bar r)}}.
\end{equation}
Express the potential \eqref{Unull} as follows:
\begin{equation}\label{b.92}
U(r) = L^2u(r),\quad u(r) := \frac{A(r)}{C(r)}.
\end{equation}

Then, relations \eqref{b.90} and \eqref{b.91}  for the spiral trajectory become:
\begin{equation} \label{b.93a}
\varepsilon = \pm |L|\sqrt{u(r_*)},
\end{equation}
\begin{equation}\label{b.93}
\phi - \phi_0 =  \pm \frac{L}{|L|}\int_{r_0}^r\frac{d\bar r}{C(\bar r)\sqrt{u(r_*) - u(\bar r)}}.
\end{equation}

To analyze the asymptotic behavior of the angular variable
as $r \to r_*$, expand $u(r)$ near $r_*$:
\begin{align}
&u(r) = u(r_*) + \frac{du}{dr}\biggr|_{r = r_*}( r - r_*) \nonumber
\\
&+\frac{1}{2} \frac{d^2u}{dr^2}\biggr|_{r = r_*} ( r - r_*)^2 + O\left(( r - r_*)^{3} \right). \label{b.94}
\end{align}
Assuming that, accordingly to \eqref{b.65}, $r_*$ is the point of maximum of the effective potential,
 we have the relations
\begin{equation}\label{b.95}
\frac{du}{dr}\biggr|_{r = r_*}\!\!\! = 0,\qquad \frac{d^2u}{dr^2} \biggr|_{r = r_*} \!\!\!= 2D < 0,
\end{equation}
from which we obtain:
\begin{equation}\label{b.96}
u(r) = u(r_*) - |D|( r - r_*)^2 + O\left((r - r_*)^3\right).
\end{equation}
Here 
\begin{equation}\label{b.96D}
D \equiv   \frac{1}{2}\frac{d^2u}{dr^2}\biggr|_{r = r_*} < 0
\end{equation}
and we deal with the case of instability of circular orbits.

Relation \eqref{b.93} describes two spiral trajectories:

(a) Outer spiral:
\begin{equation}\label{b.97}
r_* < r < r_0,
\end{equation}

(b) Inner spiral:
\begin{equation}\label{b.98}
{r}_* > r > r_0.
\end{equation}

Here, we consider the case (a), i.e. the outer spiral trajectory.
Setting (without loss of generality) $l > 0$, $\phi_0 = 0$, and choosing the ``$-$'' sign in
\eqref{b.93}, we obtain:
\begin{align}
\phi(r) = - \int_{r_0}^r\frac{d\bar r}{C(\bar r)\sqrt{u(r_*)  - u(\bar r)}}  = \int_r^{r_0}\frac{d\bar r}{C(\bar r)\sqrt{u(r_*) - u(\bar r)}}. \label{b.99}
\end{align}

Using the expansion (\ref{b.94}), we derive the asymptotic formula as
$r \to r_* + 0$ ($r_* < r < r_0$):
\begin{align}
\phi(r) \sim\int_r^{r_0}\frac{d\bar r}{C(r_*)\sqrt{|D|(\bar r - r_*)^2}} = \frac{1}{C(r_*)\sqrt{|D|}}\ln\frac{r_0 - r_*}{r - r_*}. \label{b.100}
\end{align}
This can be rewritten as:
\begin{equation}\label{b.101}
\exp\left( \beta_* \phi \right)\sim\frac{r_0 - r_*}{r - r_*}, \text{ as } r \to r_*,
\end{equation}
or, equivalently:
\begin{equation}\label{b.102}
(r_0 - r_*)\exp(- \beta_{\ast}\phi)\sim r - r_*,
\end{equation}
where parameter $\beta_*$ is defined as:
\begin{equation}\label{b.103}
\beta_{*} = C(r_*) \sqrt{|D|}.
\end{equation}

\subsection{Black hole shadows}

Now consider relation \eqref{b.99} for the special case of a spiral-like geodesic with an infinite winding angle. This curve forms a boundary between two distinct classes of solutions and corresponds to a critical angle $\vartheta_{sh} \in (0, \pi/2)$ - defined as the angle between the tangent to the curve \eqref{b.99} at the point $(r_0, \phi_0)$ and the radial line. 

If a light ray is emitted from $(r_0, \phi_0)$ at an angle smaller than this critical value:
\begin{equation}\label{b.104}
\vartheta < \vartheta_{sh},
\end{equation}
the ray will cross the photon sphere and, in the case of a black hole, proceed inward through the event horizon to fall into the singularity.

If the emission angle satisfies
\begin{equation}\label{b.105}
\vartheta > \vartheta_{sh},
\end{equation}
the ray will not reach the photon sphere and will escape to infinity after a finite number of revolutions.

Since $r_*$ is the point of maximum of the effective potential, we have:
\begin{equation}\label{b.107}
u(r_*) > u(r) > 0.
\end{equation}
for all $r > r_*$.
The \textbf{shadow
angle} $\vartheta_{sh}$ can be calculated using the
differential relation derived from \eqref{b.99},
\begin{equation}\label{b.108}
 d\phi =  - \frac{dr}{C(r)\sqrt{u(r_*) - u(r)}},
\end{equation}
and the $2D$ section of the metric (\ref{b4.m}) having the form 
\begin{equation}\label{b.109}
\!\! dl^2 = B(r)dr^2 + C(r)d\phi^2, \;\; B(r)=A(r)^{-1}.
\end{equation}

Using equations \eqref{b.108} and \eqref{b.109}, we find that:
\begin{align}
\tan\vartheta_{sh} = \frac{\sqrt{g_{\phi \phi}(r_0)}}{\sqrt{g_{rr}(r_0)}}
 \left| \frac{d\phi}{dr} \right|_{r = r_0} = \frac{1}{\sqrt{\frac{u(r_*)}{u\left( r_0 \right)} - 1}}. \label{b.110}
\end{align}
Using the trigonometric identity we get
\begin{equation}\label{b.112}
\sin^2\vartheta_{sh} = \frac{1}{1 + \cot^2\vartheta_{sh}} = \frac{u(r_0)}{u(r_*)},
\end{equation}
or, equivalently \cite{PTs}:
\begin{equation}\label{b.113}
\sin\vartheta_{sh} = \sqrt{\frac{u(r_0)}{u(r_*)}}.
\end{equation}
Therefore, Eq.~\eqref{b.107} directly gives the shadow angle::
\begin{equation}\label{b.114}
\vartheta_{sh} = \arcsin\sqrt{\frac{u(r_0)}{u(r_*)}}, \qquad 0<\vartheta_{sh}<\frac{\pi}{2},
\end{equation}
for all $r_0 > r_*$.
Here, $r_0$ describes the radial coordinate (position) of a point light source or light receiver (observer),
and $r_*$ is the radius of the photon sphere.

When the observer's radial coordinate $r_0$ is sufficiently large ($r_0 \gg M$), the shadow angle can be approximated with a certain accuracy by the asymptotic formula:
\begin{equation}\label{b.115}
   \vartheta_{sh} = \frac{b_*}{r_0 } + O \left( \frac{1}{r_0^2}  \right), 
   \end{equation}
as $r_0 \to + \infty$. Here
 \begin{equation}\label{b.116}
   b_{*} = \frac{1}{\sqrt{u(r_*)}} 
   \end{equation} 
is critical impact parameter
%:  $b_{*} = b(r_*)$, where  
 %\begin{equation}\label{b.117}
 %   b(r_0) = \frac{1}{\sqrt{u(r_0)}}, 
 %   \end{equation}
%is impact parameter function 
\cite{PTs}.   \\
\\

\textbf{Example 1: Shadow in the Schwarzschild spacetime for a static observer}

For the Schwarzschild spacetime
\begin{equation}\label{Sch_1}
   A(r) = 1 - \frac{2M}{r}, \quad C(r) = r^2,
   \end{equation}
and the function $u(r_\ast)$ and the radius of the photon sphere $r_\ast$ are defined as:
\begin{equation}\label{Sch_2}
  u(r_\ast) = \frac{A(r_\ast)}{C(r_\ast)}, \quad r_\ast = 3M.
   \end{equation}
Using equations \eqref{b.92}, \eqref{b.115} and \eqref{b.116}, we find shadow of Schwarzschild black hole for large distances:
\begin{equation}\label{Sch_3}
   \vartheta_{sh} = \frac{b_*}{r_0 } = \frac{3\sqrt{3}M}{r_0}, \quad r_0 \gg M.
   \end{equation}
\\
\\

\textbf{Example 2: Shadow in the exponential  asymptotically Schwarzschild-like spacetime 1 for a static observer}

In the exponential spacetime \eqref{1A_55} 
\begin{equation}\label{EXP_1_1}
   A(r) = e^{-\frac{2M}{r + l}}, \quad C(r) = r^2,
   \end{equation}
and we have
\begin{equation}\label{EXP_1_2}
  u(r_\ast) = \frac{A(r_\ast)}{C(r_\ast)}, \quad r_{\ast} = \frac{1}{2}\Big(M - 2l + \sqrt{M^2 - 4Ml}\Big).
   \end{equation}

For the critical value of the impact parameter we obtain:
\begin{equation}\label{EXP_1_3}
  b_\ast = r_\ast \cdot \text{exp}\Bigg(\frac{M}{r_\ast + l}\Bigg).
   \end{equation}

For the special case $l = 0$:
\begin{equation}\label{EXP_1_4}
r_\ast = M, \quad b_\ast = Me^1 \approx 2.718M.
   \end{equation}
This is the specific result for this particular metric with $l = 0$, demonstrating how the parameter $l$ modifies the geometry and the resulting shadow size from the standard Schwarzschild value of $3\sqrt{3}M$. At $l \approx 0.2362M$, the shadow of this exponential black hole coincides with the shadow of a Schwarzschild black hole.\\
\\

\textbf{Example 3: Shadow in the exponential  asymptotically Schwarzschild-like spacetime 2 for a static observer}

In the exponential spacetime \eqref{1A_666} 
\begin{equation}\label{EXP_2_1}
   A(r) = e^{\frac{2M}{l}\Big[arctan\Big(\frac{r}{l}\Big) - \frac{\pi}{2}\Big]}, \quad C(r) = r^2,
   \end{equation}
and we get
\begin{equation}\label{EXP_2_2}
  u(r_\ast) = \frac{A(r_\ast)}{C(r_\ast)}, \quad r_{\ast} = \frac{M}{2}\Bigg(1  + \sqrt{1 - \Bigg(\frac{2l}{M}\Bigg)^2}\Bigg).
   \end{equation}

The critical value of the impact parameter  is:
\begin{equation}\label{EXP_2_3}
  b_\ast = r_\ast \cdot \text{exp}\Bigg(-{\frac{M}{l}\Big[arctan\Big(\frac{r_\ast}{l}\Big) - \frac{\pi}{2}\Big]}\Bigg).
   \end{equation}
Our calculations show that the shadow of this exponential black hole does not coincide with the shadow of a Schwarzschild black hole for any values of $l$.

\section{General Relativity  interpretation for the exponential metric}

The exponential metric is a fascinating and non-trivial topic in General Relativity. While it is not a vacuum solution to the Einstein Field Equations, it has compelling physical interpretations and has been the subject of significant research, particularly in attempts to model the gravitational fields of elementary particles or as an alternative to the Schwarzschild metric. The most impontant form of the exponential metric, in spherical coordinates $(t, r, \theta, \varphi)$ and using units where $c = G = 1$, is Yilmaz-Rosen metric:
\begin{eqnarray}
ds^2 = -e^{-\frac{2M}{r}}dt^2 + e^{\frac{2M}{r}}dr^2 + e^{\frac{2M}{r}}r^2\Big(d\theta^2 + sin^2\theta d\varphi^2\Big).
 \label{8C_1} 
\end{eqnarray}
Solutions of the Einstein Field Equations in this metric has following form \cite{Simp_2} :
\begin{eqnarray}
R_{\mu\nu} = -\frac{2M^2}{r^4} diag\{0, 1, 0, 0\}_{\mu\nu} = -\frac{1}{2}\nabla_\mu\Bigg(\frac{2M}{r}\Bigg)\nabla_\nu\Bigg(\frac{2M}{r}\Bigg) = -\frac{1}{2}\nabla_\mu \phi \nabla_\nu \phi.
 \label{8C_2} 
\end{eqnarray}
Equivalently
\begin{eqnarray}
G_{\mu\nu} = -\frac{1}{2}\Bigg \{\nabla_\mu \phi \nabla_\nu \phi - \frac{1}{2}g_{\mu\nu}(g^{\rho\lambda}\nabla_\rho \phi \nabla_\lambda \phi) \Bigg \}.
 \label{8C_3} 
\end{eqnarray}
This is just the usual Einstein equation for a \textit{negative kinetic energy massless scalar field}, a “ghost” or “phantom” field.

The most robust interpretation comes from analyzing the motion of a slow test particle in a weak field.  In the weak field regime $(\frac{M}{r} \ll 1)$, we can expand the exponential functions:
\begin{eqnarray}
g_{tt} = -e^{-\frac{2M}{r}} \approx -\Bigg(1 - \frac{2M}{r} + \frac{2M^2}{r^2} -  ... \Bigg).
 \label{8C_4} 
\end{eqnarray}
In General Relativity, the Newtonian gravitational potential $\phi$ is identified from the $g_{tt}$  component as $g_{tt} = -(1+ 2\phi)$. Comparing this, we find:
\begin{eqnarray}
\phi(r) = - \frac{M}{r} + \frac{M^2}{r^2} - ... \quad   .
 \label{8C_5} 
\end{eqnarray}
The leading term, $-\frac{M}{r}$ is the standard Newtonian potential. This confirms that the exponential metric correctly reproduces Newtonian gravity in the weak-field, low-velocity limit. If we consider the exact potential derived from $g_{tt}$, it is:
\begin{eqnarray}
\phi_{exp}(r) = - \frac{M}{r}e^{- \frac{2M}{r}}. 
 \label{8C_6} 
\end{eqnarray}
For weak fields, this is approximately $-\frac{M}{r}$, but the full form is a Yukawa potential. The Yukawa potential, $\phi \sim -\frac{e^{\mu r}}{r}$ is ubiquitous in physics to describe screened interactions, where the exponential term suppresses the force at long ranges.

The exponential metric can be interpreted as describing a gravitational field where the force is not purely long-range like Newtonian gravity, but is instead screened or damped. This is a radical departure from standard GR, where gravity remains unscreened. This screening effect is often associated with the presence of a "gravitational charge" or mass distribution that effectively terminates the field, making it a candidate for modeling the gravitational field of an elementary particle (like an electron) rather than a black hole.

Thus, a certain irony emerges: despite being promoted by those who, for various reasons, reject General Relativity, the exponential metric they endorse possesses a direct, albeit exotic, interpretation within General Relativity.

% % % % % % % % % % % % % % % % % % % % % % % % %
\section{Conclusions}
% % % % % % % % % % % % % % % % % % % % % % % % %

For various reasons, we have proposed an exponential spacetime metric:
\begin{eqnarray}
ds^2 = -e^{2\alpha(r)}dt^2 - e^{-2\alpha(r)}dr^2 - r^2\Big(d\theta^2 + sin^2\theta d\varphi^2\Big).
 \label{9F_1} 
\end{eqnarray}

Although it agrees with weak-field General Relativity and Newtonian gravity, its strong-field behavior is completely different. While advocates highlight the metric's lack of horizons—correctly stating it is not a black hole—they have failed to recognize that it actually describes a traversable wormhole. This carries significant and potentially problematic implications. Replacing all known black hole candidates with traversable wormholes would therefore require a careful phenomenological investigation of this radical idea.

Despite not being a vacuum solution, the exponential metric remains interesting:

- Matching Particle Physics: The Yukawa-type potential makes it attractive for attempts to unify gravity with quantum theory or to model the gravitational field of elementary particles, where a point-like mass and a screened potential might be more appropriate than a black hole geometry.

- Regular Black Holes: It serves as an example of a "regular" metric (no event horizon) that can be used to study alternatives to singular black holes.

- Theories Beyond GR: It is an exact solution in certain alternative theories of gravity, such as Yilmaz's theory of gravitation or in some theories involving a "massive" graviton (where the graviton has a small mass, leading to a Yukawa potential).

In the framework of General Relativity, the exponential metric is best interpreted as: The spacetime geometry generated by a central mass surrounded by a specific, non-physical distribution of energy and momentum. This distribution creates a gravitational potential that is of the screened Yukawa type, differing from the pure Newtonian potential of Schwarzschild. Its key observational failures (predicting only half the light bending and one-third the perihelion precession) rule it out as a viable alternative to the Schwarzschild metric for describing the solar system. However, its mathematical elegance and connection to screened potentials keep it relevant in more speculative areas of gravitational physics.

In this article, we considered examples for exponential metrics. These metrics are asymptotically flat and, at large values of $r$, are very close to the Schwarzschild metric. It can be seen that exponential metrics do not have an event horizon and a coordinate singularity at $r=0$. As shown in our research, exponential metrics are not a vacuum solution, i.e., such metrics describe a black hole surrounded by an anisotropic fluid. We managed to determine the energy conditions only for sufficiently small and large values of the radial coordinate $r$. Furthermore, the radii of the photon sphere and the ISCO in these exponential metrics were found. Although these exponential metrics are close in their properties to the Schwarzschild metric at large distances $r$, the radii of the photon sphere and the ISCO are significantly different from the case of the Schwarzschild metric. As we know, the photon sphere radius and the ISCO for a Schwarzschild black hole are $3M$ and $6M$, respectively. In the case of the exponential metric, these quantities depend not only on the black hole's mass but also on the constant $l$. The constant $l$ in these metrics is considered as a certain physical parameter of the black hole.

The Schrodinger equation describes how quantum waves behave in a potential well, predicting discrete energy levels for atoms. The Regge-Wheeler equation describes how gravitational waves behave in the potential well around a black hole, predicting discrete vibrational modes (Quasinormal Modes) for black holes. Just as the Schrödinger equation is fundamental to atomic physics, the Regge-Wheeler equation is fundamental to black hole physics. It was the first step in moving from seeing black holes as static, exotic mathematical curiosities to understanding them as dynamic, vibrating astrophysical objects that can be observed and studied. In this article, the Regge-Wheeler equations for the exponential metric are considered. We have obtained the Regge-Wheeler potential for spin 1 and 2. Graphs of the Regge-Wheeler potentials for the exponential and Schwarzschild metrics are plotted for comparison.

The shadow of a black hole is fundamentally important because it transforms black holes from theoretical predictions into observable astronomical objects. It provides a direct test of Einstein's General Relativity in extreme gravity, allows for precise measurement of a black hole's mass, and reveals the dynamics of its accretion disk and powerful jets. Ultimately, the shadow is our most powerful tool for studying the immediate environment of an event horizon, making the invisible visible. In essence, the black hole shadow moved these objects from the realm of theoretical prediction into the realm of observational fact, providing a powerful new tool to understand the most extreme objects in the universe and the fundamental laws that govern them. In this article, a detailed analysis of the shadows of black holes is made, and then the shadows of exponential and Schwarzschild black holes are calculated based on the results of our analysis.

\section*{Acknowledgements}

I thank prof. V.D. Ivashchuk for helpful discussions and useful comments on the draft.

\section{Data Availability Statements}

No Data associated in the manuscript.

\renewcommand{\theequation}{\Alph{subsection}.\arabic{equation}}
\renewcommand{\thesection}{}
\renewcommand{\thesubsection}{\Alph{subsection}}
\setcounter{section}{0}

 \end{document}